\documentclass[11pt,psfig]{article}
\usepackage{amsfonts,amssymb,amsmath}
\usepackage{epsfig}
\usepackage{natbib}
\usepackage{bm}
\usepackage{latexsym, verbatim, lscape, graphics}
\newtheorem{condition}{Condition}[section]
\newtheorem{theorem}{Theorem}[section]

\def\bPsi{\mbox{\boldmath $\Psi$}}

\def\bS{{\bf S}}

\newcommand{\bl}{\begin{flushleft}}
\newcommand{\el}{\end{flushleft}}

\newcommand{\bc}{\begin{center}}
\newcommand{\ec}{\end{center}}

\begin{document}

\baselineskip=24pt

\title{Asymptotic Properties for Cumulative Probability Models for Continuous
  Outcomes}

\author{Chun Li$^1$, Yuqi Tian$^2$, Donglin Zeng$^3$, Bryan E. Shepherd$^2$}

\date{
  $^1$University of Southern California \\
  $^2$Vanderbilt University \\
  $^3$University of North Carolina \\[2ex]
  \today
}

\maketitle

\section*{Abstract}

Regression models for continuous outcomes often require a transformation of
the outcome, which is often specified a priori or estimated from a parametric
family.  Cumulative probability models (CPMs) nonparametrically estimate the
transformation and are thus a flexible analysis approach for continuous
outcomes.  However, it is difficult to establish asymptotic properties for
CPMs due to the potentially unbounded range of the transformation.  Here we
show asymptotic properties for CPMs when applied to slightly modified data
where the outcomes are censored at the ends.  We prove uniform consistency of
the estimated regression coefficients and the estimated transformation
function over the non-censored region, and describe their joint asymptotic
distribution.  We show with simulations that results from this censored
approach and those from the CPM on the original data are very similar when a
small fraction of data are censored.  We reanalyze a dataset of HIV-positive
patients with CPMs to illustrate and compare the approaches.

\newpage
\section{Introduction}

Regression analyses of continuous outcomes often require a transformation of
the outcome to meet modeling assumptions.  Because the correct transformation
is often unknown and it may not fall in a prespecified family, it is
desirable to estimate the transformation in a flexible way.  Transformation
models have been introduced to address this issue (Cheng et al., 1995;
Hothorn et al., 2017).  These models involve a latent intermediate variable
between the input and outcome variables and two model components, one
connecting the latent variable to the outcome variable through an unknown
transformation, and the other connecting the latent variable to the input
variables as in traditional regression models.  In semiparametric
transformation models, the first component is modeled nonparametrically while
the second parametrically.

The Cox proportional hazards model (Cox, 1972) for time-to-event outcomes is
an example of a semiparametric transformation model, in which the effects of
covariates are modeled parametrically and the baseline hazard function is
modeled nonparametrically.  More general transformation models including
proportional hazards and proportional odds models have been studied
extensively in the literature.  Among them, Zeng and Lin (2007) proposed a
general nonparametric maximum likelihood framework for right censored data,
where the cumulative hazard function is estimated as a step function with
non-negative jumps at the observed failure times.  They further established
the asymptotic properties of the resulting estimators, including consistency
and asymptotic efficiency.  However, these approaches cannot be applied
directly to study transformation models for a continuous outcome, because the
outcome has no bounds on its range and there is no clear definition of a
hazard rate function for such transformation models.  Furthermore, the
algorithms proposed in Zeng and Lin (2007) are based either on brute force
optimization, which may not guarantee convergence, or on slow expectation-maximization algorithms.

Liu et al. (2017) studied the performance of semiparametric linear
transformation models for continuous outcomes.  They showed that linear
transformation models are cumulative probability models (CPMs), and that
their nonparametric likelihood function is equivalent to a multinomial
likelihood treating the outcome variable as if it were ordered categorical
with the observed values as its categories.  This result allows us to fit
semiparametric transformation models for continuous outcomes using ordinal
regression methods.  They showed with simulations that CPMs perform well in a
wide range of scenarios.  However there is no established asymptotic theory
for the method.  One main hurdle is that the unknown transformation of the
continuous outcome variable can have an unbounded range of values, which
makes it hard to establish asymptotic properties of CPMs across the whole
range.

In this paper, we prove several asymptotic properties for CPMs when they are
applied to data that are slightly modified from the original.  Briefly, a
lower bound $L$ and an upper bound $U$ for the outcome are chosen prior to
analysis, the outcomes are censored at these bounds, and then a CPM is fit to
the censored data.  We prove that in this censored approach, the
nonparametric estimate of the underlying transformation function is uniformly
consistent in the interval $[L,U]$.  We then describe its asymptotic
distribution as well as the joint asymptotic distribution for both the
estimate of the transformation function and the estimates of the coefficients
for the input variables.  We also show that the results from this censored
approach and those from the CPM on the original data are very similar when a
small fraction of data are censored at the bounds.

\section{Method}
\subsection{Cumulative Probability Models}

Let $Y$ be the outcome of interest and $Z$ be a vector of $p$ covariates.
The semiparametric linear transformation model is
\begin{equation}
  Y=H(\beta^TZ + \epsilon),
\end{equation}
where $H$ is a transformation function assumed to be non-decreasing but
unknown otherwise, $\beta$ is a vector of coefficients, and $\epsilon$ is
independent of $Z$ and is assumed to follow a continuous distribution with
cumulative distribution function $G(\cdot)$.  An alternative expression
of model (1) is
\begin{equation}
  A(Y)=\beta^TZ + \epsilon,
\end{equation}
where $A=H^{-1}$ is the inverse of $H$.  For mathematical clarity, we assume
$H$ is left continuous and define $A(y)=\sup\{z:H(z)\le y\}$; then $A$ is
non-decreasing and right continuous.

Model (1) is equivalent to the cumulative probability model (CPM) presented
in Liu et al. (2017):
\begin{equation}
  G^{-1}\{\text{P}(Y\le y\mid Z)\} =A(y)-\beta^TZ, \text{ for any } y,
\end{equation}
where $G^{-1}(\cdot)$ serves as a link function.  One example of the
distribution for $\epsilon$ is the standard normal distribution,
$G(x)=\Phi(x)$.  In this case, the CPM becomes a normal linear model after a
transformation, which includes log-linear models and linear models with a
Box--Cox transformation as special cases.  The CPM becomes a Cox proportional
hazards model when $\epsilon$ follows the extreme value distribution, i.e.
$G(x)=1-\exp(-e^x)$, or a proportional odds model when $\epsilon$ follows
the logistic distribution, i.e. $G(x)=e^x/(1+e^x)$.

Suppose the data are \text{i.i.d.} and denoted as $(Y_i, Z_i), i=1,\ldots,n$.
Liu et al. (2017) proposed to model the transformation $A$ nonparametrically.
The corresponding likelihood is
\begin{equation*}
  \prod_{i=1}^n
  \left[ G\left\{A(Y_i)-\beta^TZ_i\right\}-G\left\{A(Y_i-)-\beta^TZ_i\right\} \right],
\end{equation*}
where $A(y-)=\lim_{t\uparrow y}A(t)$.  Since $A$ can be any non-decreasing
function, this likelihood will be maximized when the increments of $A(\cdot)$
are concentrated at the observed $Y_i$.  Liu et al. (2017) showed that this
is equivalent to treating the outcomes as if they were ordered categorical
with the observed distinct values as the ordered categories, and that the
nonparametric maximum likelihood estimates can be obtained by
fitting an ordinal regression model.  They showed in simulations that CPMs
perform well under a wide range of scenarios.  However, since some observed
$Y_i$ can be extremely large or small and the observations at both tails are
often sparse, there is high variability in the estimate of $A$ at the tails.
Moreover, the unboundness of the transformation at the tails makes it
difficult to control the compactness of the estimator of $A$, thus making
most of asymptotic theory no longer applicable.

\subsection{Cumulative Probability Models on Censored Data}

In view of the challenges above, we hereby describe an approach in which the
outcomes are censored at a lower bound and an upper bound before a CPM is
fit.  We will then describe the asymptotic properties of this approach in
Section 2.3, and show with simulations that the results from this approach
and those of the CPM on the original data are similar when a small fraction
of data are censored.

More specifically, we predetermine a lower bound $L$ and an upper bound $U$,
and consider any observation outside the interval $(L,U)$ as censored.  In
other words, those with $Y_i\le L$ are treated as left-censored at $L$, and
those with $Y_i\ge U$ are treated as right-censored at $U$.  The censored
data may be denoted as
\begin{equation*}
  (Y_i', Z_i, I(Y_i\le L), I(Y_i\ge U))=
  \left\{
    \begin{array}{ll}
      (Y_i, Z_i, 0, 0), & Y_i\in(L,U); \\
      (L, Z_i, 1, 0), & Y_i\le L; \\
      (U, Z_i, 0, 1), & Y_i\ge U;
    \end{array}
  \right. \quad i=1,\ldots,n.
\end{equation*}
The bounds $L$ and $U$ should satisfy $\text{P}(L<Y<U)>0$,
$\text{P}(Y\le L)>0$, and $\text{P}(Y\ge U)>0$.  The variable $Y_i'$
follows a mixture distribution.  When $Y_i'\in (L,U)$, the distribution is
continuous with the same cumulative distribution function as that for $Y_i$;
that is,
$\text{P}(Y_i'\le y\mid Z_i) =\text{P}(Y_i\le y\mid Z_i)
=G\{A(y)-\beta^TZ_i\}$ for $y\in(L,U)$.  When $Y_i'=L$ or $Y_i'=U$, the
distribution is discrete, with
$\text{P}(Y_i'=L\mid Z_i)=G\{A(L)-\beta^TZ_i\}$ and
$\text{P}(Y_i'=U\mid Z_i)=1-G\{A(U{-})-\beta^TZ_i\}$.  Then the
nonparametric likelihood for the censored data is
\begin{align}
  \prod_{i=1}^n
  &\left(
    \left[ G\{A(Y_i)-\beta^TZ_i\}-G\{A(Y_i-)-\beta^TZ_i\}
    \right]^{I(Y_i\in(L,U))}  \right. \times \nonumber \\
  & \quad\left. G\{A(L)-\beta^TZ_i\}^{I(Y_i\le L)} \times
    \left[ 1-G\{A(U-)-\beta^TZ_i\} \right]^{I(Y_i\ge U)} \right).
\end{align}

Since $A(\cdot)$ can be any non-decreasing function over the interval
$[L,U)$, the likelihood (4) will be maximized when the increments of
$A(\cdot)$ are concentrated at the observed $Y_i'$.  Hence it suffices to
consider only step functions with a jump at each distinct value of
$Y_i'\in[L,U]$.  
%Let $Y_{\text{min}}' =\min\{Y_i':Y_i>L\}$ and
%$Y_{\text{max}}' =\max\{Y_i':Y_i<U\}$.  On the lower end, if there are
%observations with $Y_i'=L$, the second term in the product in (4) comes into
%play, and any function $A(\cdot)$ that maximizes (4) will have a finite
%$A(L)$ and a finite jump size at $Y_{\text{min}}'$.  If there are no
%observations with $Y_i'=L$, the likelihood will be maximized when
%$A(Y_{\text{min}}'-)=-\infty$ (i.e., an infinite ``jump size'' at
%$Y_{\text{min}}'$).  On the upper end, if there are observations with
%$Y_i'=U$, the third term in the product in (4) comes into play, and any
%function $A(\cdot)$ that maximizes (4) will have a finite jump size at
%$Y_{\text{max}}'$ and $A(U-)=A(Y_{\text{max}}')$.  If there are no
%observations with $Y_i'=U$, the likelihood will be maximized when
%$A(Y_{\text{max}}')=\infty$ (i.e., an infinite ``jump size'' at
%$Y_{\text{max}}'$).

\subsection{Asymptotic Results}

From now on we assume the outcome is continuous.  Without loss of generality,
we assume that in our models (1)--(3), the support of $Y$ contains $0$, the
vector $Z$ contains an intercept and has $p$ dimensions, and $A(0)=0$.
Furthermore, the bounds for censoring satisfy $L<0$ and $U>0$.  To establish
the asymptotic properties described below, we further assume
\begin{condition}
  $G(x)$ is thrice-continuously differentiable, $G'(x)>0$ for any $x$,
  $G''(x)\textrm{sign}(x)<0$ for $\vert x \vert \ge M$, where $M>0$ is a
  constant, and
  $$\liminf_{x\rightarrow \infty}G'(x)/\{1-G(x)\}>0, \ \ \liminf_{x\rightarrow
    -\infty}G'(x)/G(x)>0.$$
\end{condition}
\begin{condition}
The covariance matrix of $Z$ is non-singular.    
In addition, $Z$ and $\beta$ are
  bounded so that $\beta^TZ\in[-m, m]$ almost surely for some large constant
  $m$.
\end{condition}
\begin{condition}
$A(y)$ is continuously differentiable in $(-\infty,\infty)$.
\end{condition}
Condition 1 imposes restrictions on $G(x)$ at both tails; it holds for
many residual distributions, including the standard normal distribution, the
extreme value distribution and the logistic distribution. Conditions 2
and 3 are minimal assumptions for establishing asymptotic properties for
linear transformation models.

Let $(\widehat\beta, \widehat A)$ denote the nonparametric maximum likelihood
estimate of $(\beta, A)$ that maximizes the likelihood (4) of the censored
approach described in Section 2.2.  Then $\widehat A$ is a step function with
a jump at each of the distinct $Y_i'$ in the censored data.  To establish the
asymptotic properties for $(\widehat\beta, \widehat A)$, we consider
$\widehat A$ as a function over the closed interval $[L,U]$ by defining
$\widehat A(U)=\widehat A(U-)$.  We have the following consistency theorem.

\begin{theorem}
  Under conditions 1 -- 3, with probability one,
  $$\sup_{y\in [L, U]}\vert \widehat A(y)-A(y)\vert
  +\Vert\widehat\beta-\beta\Vert\rightarrow 0.$$
\end{theorem}
%Theorem 1 holds for any $L$ and $U$ that satisfy $\text{P}(L<Y<U)>0$, $\text{P}(Y\le L)>0$, and $\text{P}(Y\ge U)>0$. 
The proof of Theorem 1 is in Supplementary material.  Core steps of the proof
include showing that $\widehat A$ is bounded in $[L,U]$ with probability one.
Then, since $\widehat A(\cdot)$ is bounded and increasing in $[L, U]$, by the
Helly selection theorem, for any subsequence, there exists a weakly
convergent subsequence.  Thus, without confusion, we assume that
$\widehat A\rightarrow A^*$ weakly in $[L, U]$ and
$\widehat\beta \rightarrow\beta^*$.  We then show that with probability one,
$A^*(y)=A(y)$ for $y\in [L, U]$ and $\beta^*=\beta$.  With this result, the
consistency is established.  Furthermore, since $A$ is continuously
differentiable, we conclude that $\widehat A(y)$ converges to $A(y)$
uniformly in $[L, U]$ with probability one.

We next describe the asymptotic distribution for
$(\widehat\beta, \widehat A)$.  The asymptotic distribution of $\widehat A$
will be expressed as that of a random functional in a metric space.  We first
define some notation.  Let $BV[L,U]$ be the set of all functions defined over
$[L,U]$ for which the total variation is at most one.  Let $lin(BV[L,U])$ be
the set of all linear functionals over $BV[L,U]$; that is, every element $f$
in $lin(BV[L,U])$ is a linear function $f: BV[L,U]\rightarrow R$.  For any
$f\in lin(BV[L,U])$, its norm is defined as
$\Vert f\Vert =\sup_{h\in BV[L,U]}f[h]$.  A metric over $lin(BV[L,U])$ can
then be derived subsequently.  Given any non-decreasing function $A$ over
$[L,U]$, a corresponding linear functional in $lin(BV[L,U])$, also denoted as
$A$, can be defined such that for any $h\in BV[L,U]$,
$$A[h]=\int_L^U h(x)dA(x).$$
Similarly, for an nonparametric maximum likelihood estimate $\widehat A$, its corresponding linear functional in
$lin(BV[L,U])$ is $\widehat A$ such that for any $h\in BV[L,U]$,
$$\widehat A[h]=\int_L^U h(x)d\widehat A(x).$$
The functional $\widehat A$ is a random element in the metric space
$lin(BV[L,U])$.
% The next theorem gives the asymptotic distribution for
% $n^{1/2} (\widehat\beta-\beta, \widehat A-A)$.
For any $y\in(L,U)$, there exists an $h\in BV[L,U]$ such that
$\widehat A(y) =\widehat A[h]$.  For example, suppose the estimated jump
sizes at the distinct outcome values of a dataset, $\{a_1,\ldots,a_J\}$, are
$\{\hat s_1,\ldots,\hat s_J\}$.  Then at $y_0>0$,
$\widehat A(y_0) =\sum_{0<a_j\le y_0} \hat s_j =\widehat A[h_0]$, where
$h_0(y)=I(0<y\le y_0)$; and similarly, at $y_0<0$,
$\widehat A(y_0) =-\sum_{y_0<a_j<0} \hat s_j =\widehat A[h_0]$, where
$h_0(y)=I(y_0<y<0)$.

\begin{theorem}
Under conditions 1 -- 3, 
$n^{1/2} (\widehat\beta-\beta, \widehat A-A)$ converges weakly to a tight
Gaussian process in $R^p\times lin(BV[L, U])$.  Furthermore, the asymptotic
variance of $n^{1/2} (\widehat\beta-\beta)$ attains the semiparametric
efficiency bound.
\end{theorem}

The proof of Theorem 2 is in the Supplementary material and makes use modern empirical process and semiparametric efficiency theory.  Its proof relies on verifying all the technical conditions in the
Master Z-Theorem in van der Vaart and Wellner (1996).  In particular, it
entails verification of the invertibility of the information operator for
$(\beta, A)$.  %The details of the proofs are given in the Appendix.

Because the information operator for $(\beta,A)$ is invertible, the arguments
given in Murphy and van der Vaart (2000) and Zeng and Lin (2006) imply that
the asymptotic variance-covariance matrix of
$(\widehat\beta, \widehat A[h_1], \ldots, \widehat A[h_m])$ for any
$h_1,\ldots,h_m\in BV[L,U]$ can be consistently estimated based on the
information matrix for $\widehat\beta$ and the jump sizes of $\widehat A$.
Specifically, suppose the estimated jump sizes at the distinct outcome values
of a dataset, $\{a_1,\ldots,a_J\}$, are $\{\hat s_1,\ldots,\hat s_J\}$.  Let
$\widehat I_n$ be the estimated information matrix for both $\widehat\beta$
and $\{\hat s_1,\ldots,\hat s_J\}$.  Then the variance-covariance matrix for
$(\widehat\beta, \widehat A[h_1], \ldots, \widehat A[h_m])$ is estimated as
$V^T\widehat I_n^{-1} V$, where
$$V=\left(\begin{matrix} I_{p\times p} & 0 \cr 0 & H\end{matrix}\right)$$
and $H$ is a $J\times m$ matrix with elements
$\{h_k(a_j)\}_{1\le j\le J, 1\le k\le m}$.

\section{Simulation Study}

CPMs have been extensively simulated elsewhere to justify their use, and have
been largely seen to have good behavior (Liu et al., 2017; Tian et al.,
2020). Here we perform a more limited set of simulations to illustrate three
major points which are particularly relevant for our study:
% when properly specified with moderate sample sizes and to be fairly robust
% to slight model misspecifications
\begin{enumerate}
\item Estimation of $A(y)$ using CPMs can be biased at extreme values of $y$.
  Even though $\widehat A(y)$ may be consistent point-wise for any $y$,
  $\widehat A(y)$ may not be uniformly consistent over all
  $y\in(-\infty,\infty)$.
\item In the censored approach, $\widehat A(y)$ is uniformly consistent over
  $y\in(L,U)$.
  % (``well-behaved'' in finite samples?)  This point demonstrates that the
  % large sample theory derived in this study is applicable with moderate
  % sample sizes.
\item Except for estimation of extreme quantiles and $A(y)$ at extreme
  levels, results are largely similar between the uncensored and the censored
  approaches.
  % This point suggests that as a default with continuous data, one could choose arbitrary values of L and U, apply censored CPMs, and be confident that subsequent results (which have the backing of asymptotic theory) will be very similar to those had one simply applied CPMs to uncensored data (which does not have the large sample theory to back it up). In contrast, these results may also be used to justify the use of CPMs to continuous data without censoring because of their similarity to results from censored CPMs.
\end{enumerate}

\subsection{Simulation Set-up}

CPMs have been extensively simulated elsewhere to justify their use, and have
been largely seen to have good behavior (Liu et al., 2017; Tian et al.,
2020). Here we perform a more limited set of simulations to illustrate three
major points which are particularly relevant for our study: First, estimation of $A(y)$ using CPMs can be biased at extreme values of $y$.
  Even though $\widehat A(y)$ may be consistent point-wise for any $y$,
  $\widehat A(y)$ may not be uniformly consistent over all
  $y\in(-\infty,\infty)$.
Second, in the censored approach, $\widehat A(y)$ is uniformly consistent over
  $y\in(L,U)$.
  Third, except for estimation of extreme quantiles and $A(y)$ at extreme
  levels, results are largely similar between the uncensored and the censored
  approaches.

We roughly followed the simulation settings of Liu et al. (2017).  Let
$X_1 \sim$ Bernoulli(0.5), $X_2 \sim N(0,1)$, and
$Y=\exp(\beta_1 X_1 +\beta_2 X_2 + \epsilon)$, where $\beta_1=1$,
$\beta_2=-0.5$, and $\epsilon \sim N(0,1)$.  In this set-up, the correct
transformation function is $A(y)=\log(y)$.  We generated datasets
$\{(X_1,X_2,Y)\}$ with sample sizes $n=100$, 1000, and 5000.  We fit CPMs
that have correctly specified link function (probit) and model form
(linear). (Performance of misspecified models was extensively studied via
simulations in Liu et al., 2017.)  In CPMs, the transformation $A$ and the
parameters $(\beta_1, \beta_2)$ were semi-parametrically estimated.  We
evaluated how well the transformation was estimated by comparing
$\widehat A(y)$ with the correct transformation, $A(y)=\log(y)$, for various
values of $y$.

We fit CPMs on the original data without censoring and CPMs on the censored
data with censoring at $L$ and $U$, with $[L,U]$ being set to be
$[e^{-4},e^4]$, $[e^{-2},e^2]$, and $[e^{-1/2},e^{1/2}]$; these values
correspond to approximately $0.2\%, 13\%,$ and $71\%$ of $Y$ being censored,
respectively.  All simulations had $1000$ replications.

\subsection{Simulation Results}

\begin{figure}
  \begin{center}
    \includegraphics[width=5.5in]{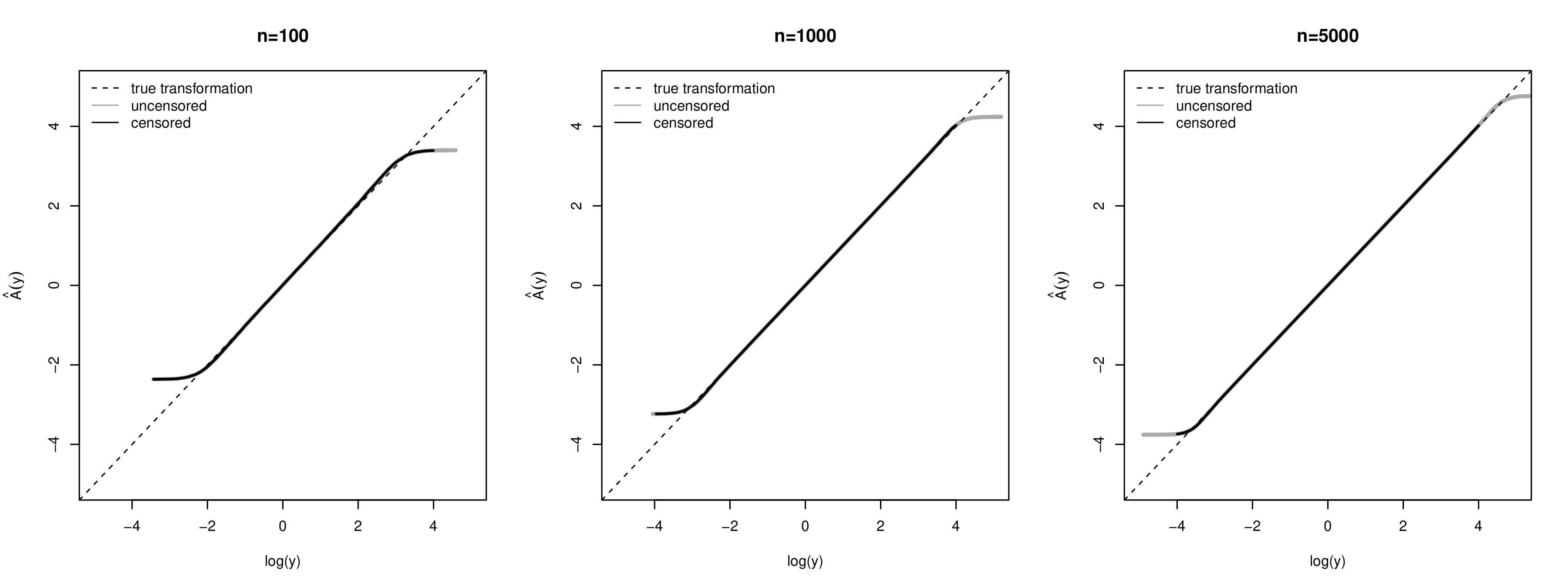}
    \includegraphics[width=5.5in]{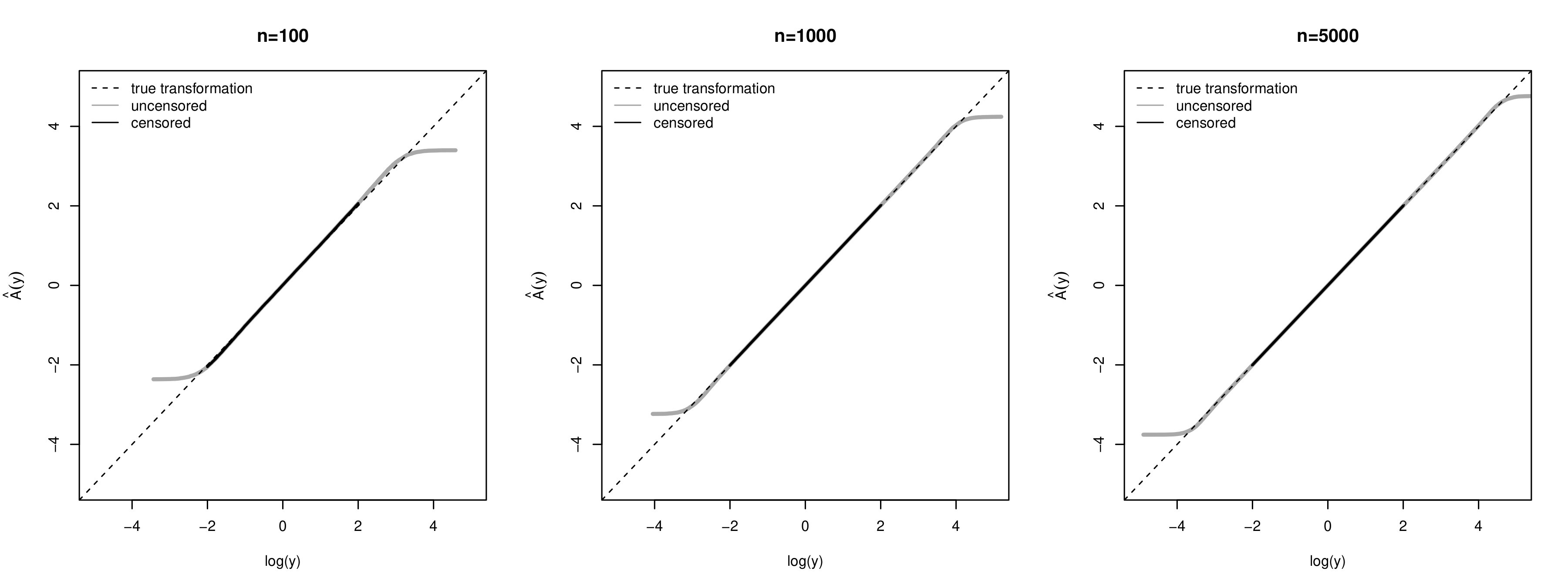}
    \includegraphics[width=5.5in]{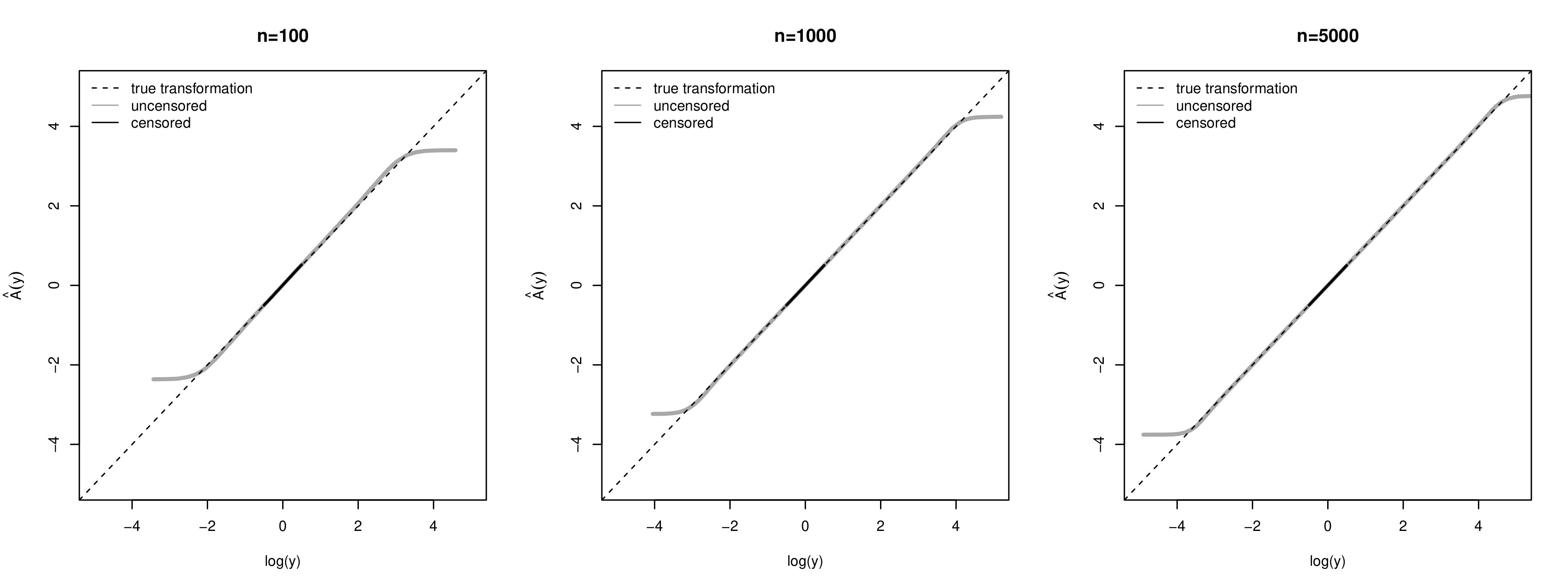}
  \end{center}
  \caption{Average estimate of $A(y)$ after fitting properly specified
    CPMs compared to the true transformation,
    $\log(y)$.  Gray curve: uncensored data; black curve: censored data.
    Dashed lines are the diagonal.  Top row: $[L,U]=[e^{-4},e^{4}]$; middle
    row: $[L,U]=[e^{-2},e^{2}]$; bottom row: $[L,U]=[e^{-1/2},e^{1/2}]$.
    Left to right: $n=100,1000,5000$.  Based on 1000 replicates.}
  \label{fig:alpha_y}
\end{figure}

\begin{figure}
  \begin{center}
    \includegraphics[width=5.5in]{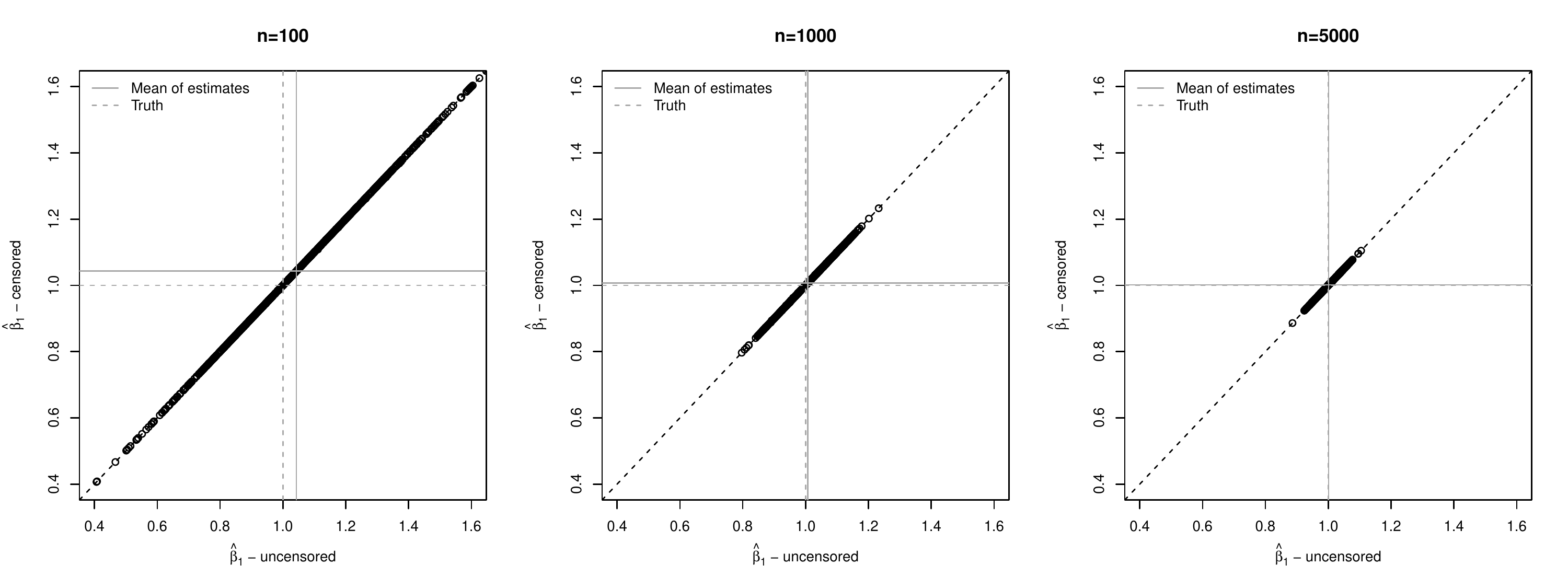}
    \includegraphics[width=5.5in]{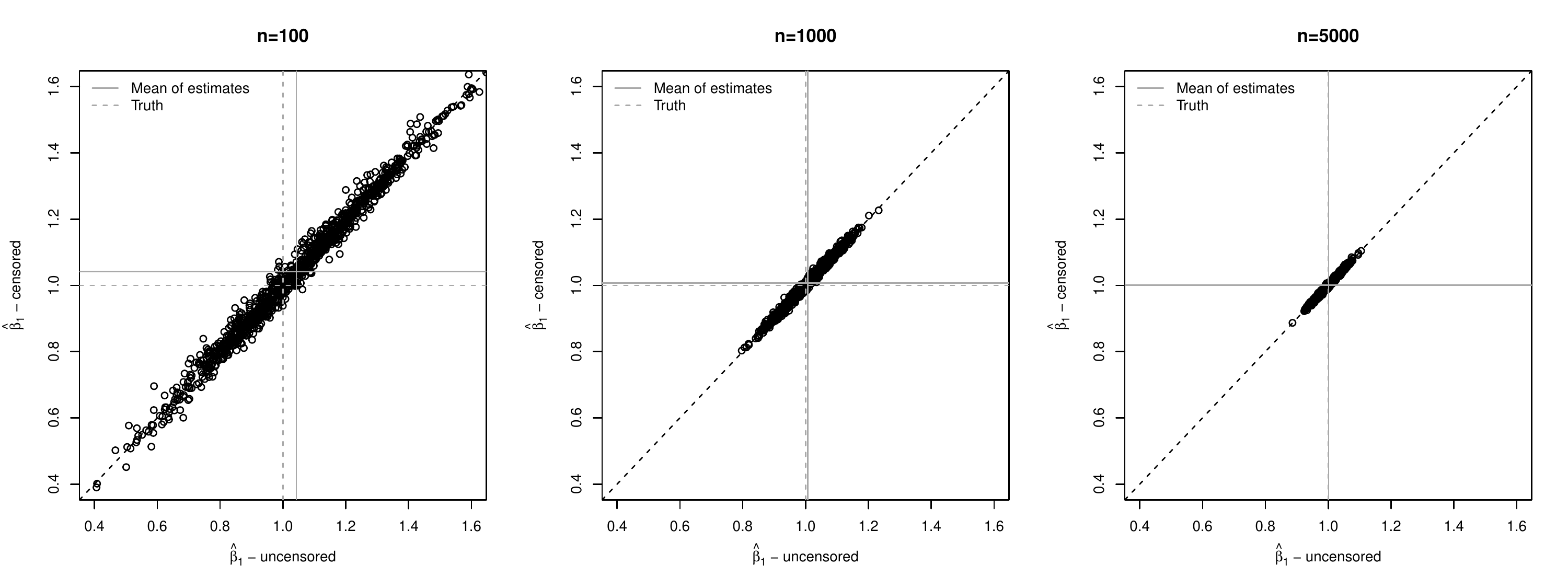}
    \includegraphics[width=5.5in]{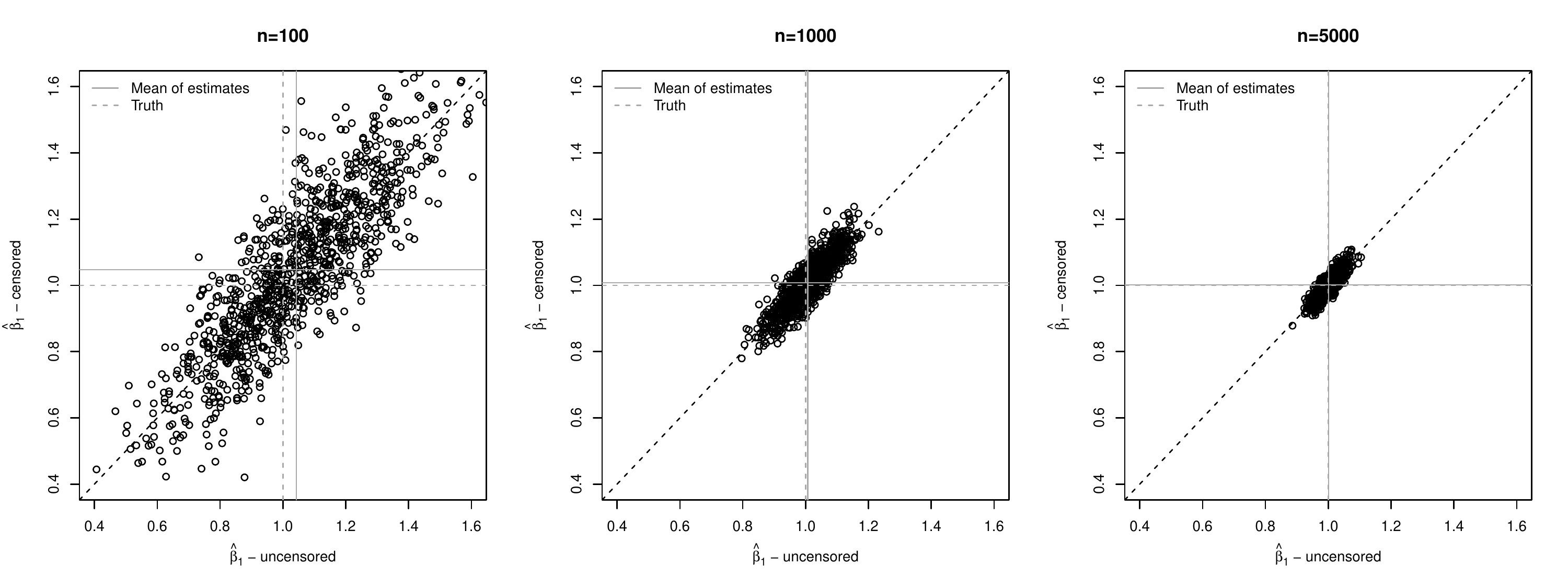}
  \end{center}
  \caption{Estimates of $\beta_1$ using data censored at $[L,U]$ compared to
    those using uncensored data and to the truth, $\beta_1=1$.  Gray lines
    are mean estimates and dashed gray lines are the truth.  Top row:
    $[L,U]=[e^{-4},e^{4}]$; middle row: $[L,U]=[e^{-2},e^{2}]$; bottom row:
    $[L,U]=[e^{-1/2},e^{1/2}]$.  Left to right: $n=100,1000,5000$.  Based on
    1000 replicates.}
  \label{fig:beta1}
\end{figure}

Figure \ref{fig:alpha_y} shows the average estimate of $A(y)$ across 1000
simulation replicates compared with the true transformation, $\log(y)$.  The
left, center, and right panels are results based on sample sizes of 100,
1000, and 5000, respectively.  With uncensored data, for all sample sizes,
estimates are unbiased when $y$ is in the center of the distribution,
approximately in the range $[e^{-2}, e^{3}]$ when $n=100$, in
$[e^{-3}, e^{4}]$ when $n=1000$, and in a wider range when $n=5000$.
However, at extreme values of $y$ we see biased estimation.
This illustrates that for a fixed $y$, one can find a sample size large
enough so that estimation of $A(y)$ is unbiased, but that there will always
be a more extreme value of $y$ for which $\widehat A(y)$ may be biased.  This
motivates the need to censor values outside a predetermined range $[L,U]$ to
achieve uniform consistency of $\widehat A(y)$ for $y\in[L,U]$.

Figure \ref{fig:beta1} compares estimates of $\beta_1$ for the various sample
sizes using uncensored data and using data censored at the three ranges of
$[L,U]$.  As sample size becomes larger, $\hat\beta_1$ becomes less biased in
all the approaches.  At $n=5000$, $\hat\beta_1$ is approximately unbiased
even with severely censored data.  Not surprisingly, with increasing levels
of censoring, $\hat\beta_1$ becomes slightly more variable (Table
\ref{tab:sim}) and slightly less correlated with that estimated from
uncensored data.  The results for $\beta_2$ have similar patterns (Supplementary material Fig. 1).

\begin{table}[!tbp]
  \caption{Simulation results for estimates from CPMs on uncensored data and those on data censored at $[L,U]$; based on
    1,000 replicates}
  \begin{tabular}{lllcccc}
    \hline
    $n$ & Estimand & & Uncensored & \multicolumn{3}{c}{Data censored at $[L,U]$} \\
        &&& Data & $[e^{-4},e^4]$ & $[e^{-2},e^2]$ & $[e^{-1/2},e^{1/2}]$  \\
    \hline
    $100$ & $\beta_1$
                   & bias    & 0.043 & 0.043 & 0.042 & 0.048\\
        &          & SD      & 0.228 & 0.228 & 0.229 & 0.260\\
        &          & mean SE & 0.217 & 0.217 & 0.219 & 0.251\\
        &          & MSE     & 0.054 & 0.054 & 0.054 & 0.070\\
        & $\beta_2$ & bias   & --0.022 & --0.021 & --0.020 & --0.022\\
        &          & SD      & 0.119 & 0.119 & 0.120 & 0.143\\
        &          & mean SE & 0.110 & 0.110 & 0.111 & 0.133\\
        &          & MSE     & 0.015 & 0.015 & 0.015 & 0.021\\
        & $A(e^{0.5})$ & bias & 0.019 & 0.019 & 0.019 & 0.020\\
        &          & SD      & 0.177 & 0.177 & 0.177 & 0.183\\
        &          & mean SE & 0.174 & 0.174 & 0.175 & 0.182\\
        &          & MSE     & 0.032 & 0.032 & 0.032 & 0.034\\
        & $\text{median}(Y \mid X_1=0, X_2=0)$
                   & bias    & 0.022 & 0.022 & 0.023 & 0.021\\
        &          & SD      & 0.172 & 0.172 & 0.172 & 0.176\\
%        &          & mean SE & - & - & - & - \\
        &          & MSE     & 0.030 & 0.030 & 0.030 & 0.031\\
        & $E(Y \mid X_1=0, X_2=0)$
                   & bias    & --0.007 & - & - & - \\
        &          & SD      & 0.266 & - & - & - \\
        &          & mean SE & 0.262 & - & - & - \\
        &          & MSE     & 0.071 & - & - & - \\
    \\
    $1000$ & $\beta_1$
                   & bias    & 0.007 & 0.007 & 0.007 & 0.008\\
        &          & SD      & 0.068 & 0.068 & 0.068 & 0.076\\
        &          & mean SE & 0.067 & 0.067 & 0.068 & 0.077\\
        &          & MSE     & 0.005 & 0.005 & 0.005 & 0.006\\
        & $\beta_2$ & bias   & --0.001 & --0.001 & --0.001 & --0.001\\
        &          & SD      & 0.033 & 0.033 & 0.034 & 0.040\\
        &          & mean SE & 0.034 & 0.034 & 0.034 & 0.041\\
        &          & MSE     & 0.001 & 0.001 & 0.001 & 0.002\\
        & $A(e^{0.5})$ & bias & 0.003 & 0.003 & 0.003 & 0.003\\
        &          & SD      & 0.055 & 0.055 & 0.055 & 0.056\\
        &          & mean SE & 0.054 & 0.054 & 0.054 & 0.057\\
        &          & MSE     & 0.003 & 0.003 & 0.003 & 0.003\\
        & $\text{median}(Y \mid X_1=0, X_2=0)$
                   & bias    & 0.003 & 0.003 & 0.002 & 0.002\\
        &          & SD      & 0.054 & 0.054 & 0.054 & 0.056\\
%        &          & mean SE & - & - & - & - \\
        &          & MSE     & 0.003 & 0.003 & 0.003  & 0.003\\
        & $E(Y \mid X_1=0, X_2=0)$
                   & bias    & --0.003 & - & - & - \\
        &          & SD      & 0.081 & - & - & - \\
        &          & mean SE & 0.083 & - & - & - \\
        &          & MSE     & 0.007 & - & - & - \\
    \hline
%    $5000$ & $\beta_1$
%                   & bias    & 0.0014 & 0.0014 & 0.0012 & 0.0017\\
%        &          & SD      & 0.0302 & 0.0302 & 0.0304 & 0.0352\\
%        &          & mean SE & 0.0300 & 0.0300 & 0.0302 & 0.0345\\
%        &          & MSE     & 0.0009 & 0.0009 & 0.0009 & 0.0012\\
%        & $\beta_2$ & bias   & -0.0006 & -0.0006 & -0.0004 & -0.0009 \\
%        &          & SD      & 0.0154 & 0.0154 & 0.0157 & 0.0189\\
%        &          & mean SE & 0.0150 & 0.0150 & 0.0153 & 0.0181\\
%        &          & MSE     & 0.0002 & 0.0002 & 0.0002 & 0.0004\\
%        & $A(e^{0.5})$ & bias & 0.0019 & 0.0019 & 0.0018 & 0.0019\\
%        &          & SD      & 0.0248 & 0.0248 & 0.0249 & 0.0260\\
%        &          & mean SE & 0.0242 & 0.0242 & 0.0243 & 0.0253\\
%        &          & MSE     & 0.0006 & 0.0006 & 0.0006 & 0.0007\\
%        & $med(Y \mid X_1=0, X_2=0)$
%                   & bias    & -0.0005 & -0.0005 & -0.0004 & -0.0005\\
%        &          & SD      & 0.0245 & 0.0245 & 0.0246 & 0.0255\\
% %       &          & mean SE & - & - & - & - \\
%        &          & MSE     & 0.0006 & 0.0006 & 0.0006 & 0.0007\\
%        & $E(Y \mid X_1=0, X_2=0)$
%                   & bias    & -0.0016 & - & - & - \\
%        &          & SD      & 0.0372 & - & - & - \\
%        &          & mean SE & 0.0370 & - & - & - \\
%        &          & MSE     & 0.0014 & - & - & - \\
  \end{tabular}
  \label{tab:sim}
%  \begin{tabnotes}
%    SD, standard deviation of replicates; mean SE, average estimated standard error across replicates; MSE, mean squared error.
%  \end{tabnotes}
\end{table}

Table \ref{tab:sim} shows further results for five estimands: $\beta_1$,
$\beta_2$, $A(e^{0.5})$, and the conditional median and mean of $Y$ given
$X_1=0$ and $X_2=0$.  For each estimand, we compute the bias of the
corresponding estimate, its standard deviation across replicates, mean of
estimated standard errors, and mean squared error.  For the estimands
$\beta_1$, $\beta_2$, and $A(e^{0.5})$, estimation using uncensored data
appears to be consistent, and the behavior of our estimators in the censored
approach is as expected by the asymptotic theory.  When $n=100$ there appears
to be only a modest amount of bias, even with 71\% censoring; when $n=1000$
and 5000 (shown in Supplementary material), bias is quite small.  Although in
Fig.  \ref{fig:alpha_y} we saw that estimates of $A(y)$ for extreme values of
$y$ were biased, we see no evidence that this impacts the estimation of
$\beta_1$ and $\beta_2$.  The average standard errors are very similar to the
empirical results (i.e., the standard deviation of parameter estimates across
replicates), suggesting that we are correctly estimating standard errors.
These results hold regardless of the amount of censoring in our simulations.
With increasing levels of censoring, as expected, both absolute bias and
standard deviation increase, and as a result, the mean squared error
increases.  However all these measures become smaller as the sample size
increases.

We cannot compute the standard error for conditional median.
Censoring also prohibits sound estimation of conditional mean; while one
could instead estimate the trimmed conditional mean, e.g.,
$E(Y \mid X_1=0, X_2=0, L \leq Y \leq U)$, which may substantially differ from
$E(Y \mid X_1=0, X_2=0)$.  The bias of $\widehat A(y)$ for extreme
values of $y$ had little impact on the estimation of $E(Y \mid X_1=0, X_2=0)$,
which is computed using $\widehat A(y)$ over the entire range of observed
$y$.

\section{Example Data Analysis}

\begin{figure}
  \begin{center}
    \includegraphics[width=5.5in]{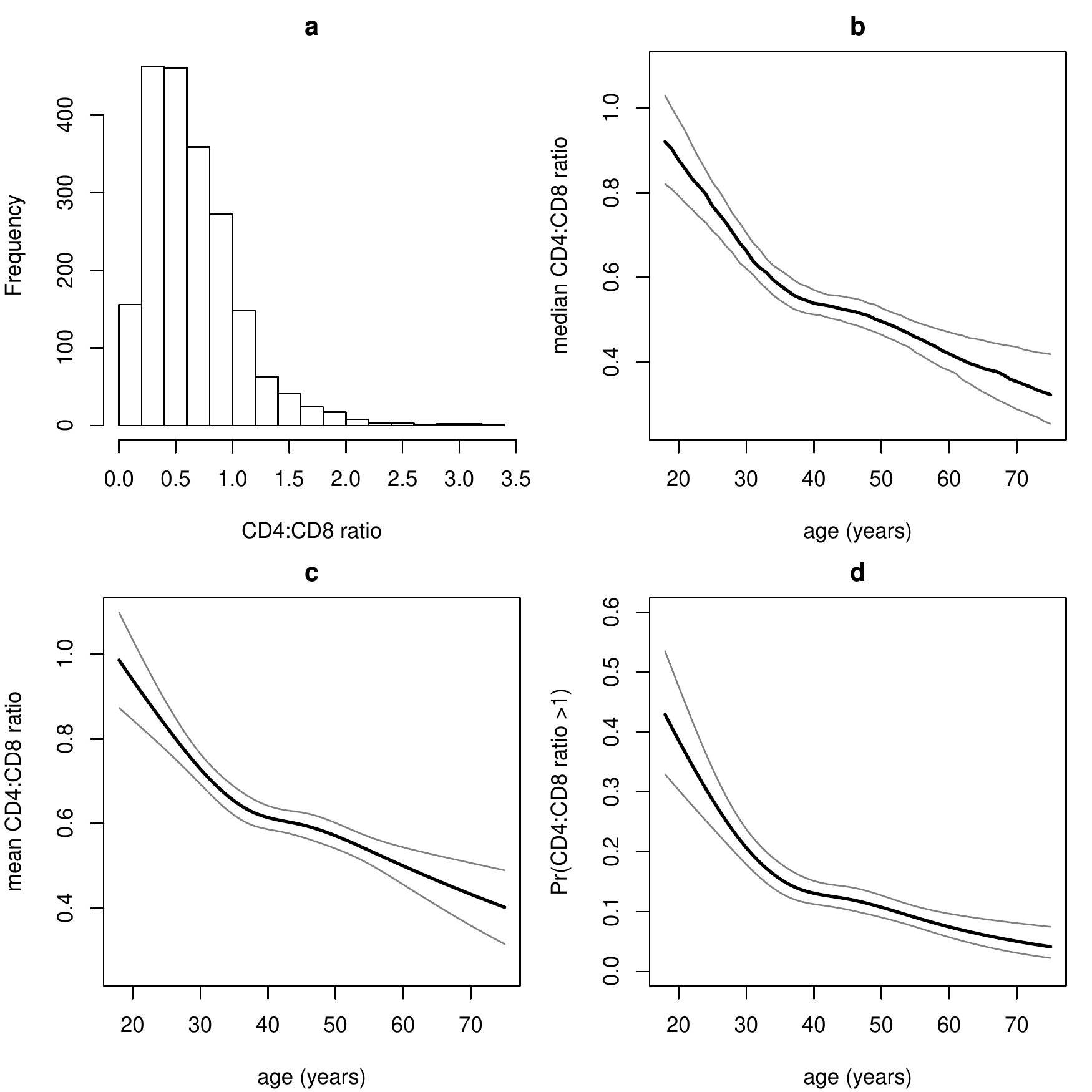}
  \end{center}
  \caption{a: Histogram of CD4:CD8 ratio in our dataset.  b--d: Estimated
    outcome measures and 95\% confidence intervals as functions of age, holding other covariates constant
    at their medians/modes.  b: median CD4:CD8 ratio; c: mean CD4:CD8 ratio;
    d: probability that CD4:CD8 $>1$.}
\end{figure}

CD4:CD8 ratio is a biomarker to measure the strength of the immune system.  A
normal CD4:CD8 ratio is between 1 and 4, while people with HIV tend to have
much lower values, and a low CD4:CD8 ratio is highly predictive of poor
outcomes including non-communicable diseases and mortality. When people with
HIV are put on antiretroviral therapy, their CD4:CD8 ratio tends to increase,
albeit often slowly and quite variably.  Castilho et al. (2016) studied
factors associated with CD4:CD8 ratio among 2,024 people with HIV who started
antiretroviral therapy and maintained viral suppression for at least 12
months.  They considered various factors including age, sex, race, probable
route of transmission, hepatitis C co-infection, hepatitis B co-infection,
and year of antiretroviral therapy initiation.  Here we re-analyze their data
using CPMs.  We will focus on the associations of
CD4:CD8 ratio with age and sex, treating the other factors as covariates.
CD4:CD8 ratio tends to be right skewed (Fig. 3a), but there is no standard
transformation for analyzing it.  In various studies, it has been
untransformed (Castilho et al., 2016), log-transformed (Sauter et al., 2016),
dichotomized (CD4:CD8 $>1$ vs.  $\le1$; Petoumenos et al., 2017), put into
ordered categories roughly based on quantiles (Serrano-Villar et al., 2015),
square-root transformed (Silva et al., 2018), and fifth-root transformed
(Gras et al., 2019).  In contrast, CPMs do not
require specifying the transformation.

We fit three CPMs: Model 1 using the original data, Model 2 censoring all
CD4:CD8 ratios below $L=0.1$ and above $U=2.0$, and Model 3 censoring below
$L=0.2$ and above $U=1.5$.  In a similar group of patients in a prior study
(Serrano-Villar et al., 2014), these values of $L$ and $U$ were approximately
the 1.5th and 99.5th percentiles, respectively, for Model 2, and the 7th and
95th percentiles for Model 3.  In our dataset, there were 19 (0.9\%) CD4:CD8
ratios below $0.1$ and 21 (1\%) above $2.0$, and 156 (7.7\%) below $0.2$ and
74 (3.7\%) above $1.5$.  In our models, age was modeled using restricted
cubic splines with four knots at the 0.05, 0.35, 0.65, and 0.95 quantiles.
All models were fit using a logit link function; quantile-quantile plots of
probability-scale residuals (Shepherd et al., 2016) from the models suggested
good model fit (Supplementary material Fig. 2).

All three models produced nearly identical results.  Female sex had
regression coefficients $0.6002$, $0.6000$, and $0.5994$ in Models 1, 2, and
3, respectively (likelihood ratio $p<0.0001$ in all models), suggesting that
the odds of having a higher CD4:CD8 ratio, after controlling for all other
variables in the model, were about $e^{0.6}=1.82$ times higher for females
than for males (95\% Wald confidence interval 1.44--2.31).  The median
CD4:CD8 ratio holding all other covariates fixed at their medians/modes was
estimated to be 0.67 (0.60--0.74) for females compared to 0.53 (0.51--0.56)
for males; all models had the same estimates to two decimal places.  The mean
CD4:CD8 ratio holding all other covariates constant was estimated to be 0.73
(0.67--0.79) for females and 0.61 (0.58--0.63) for males from Model 1.  The
mean estimates from Models 2 and 3 were slightly different (e.g., 0.72 for
females); however, the mean should not be reported after censoring because
the estimates arbitrarily assigned the censored values to be $L$ and $U$.

Older age was strongly associated with a lower CD4:CD8 ratio ($p<0.0001$ in
all models), and the association was non-linear ($p=0.0080, 0.0081, 0.0086$,
respectively).  Fig. 3b--d show the estimated median and mean CD4:CD8 ratio
and the probability that CD4:CD8 $>1$ as functions of age, all derived from
the CPMs and holding other covariates fixed at their
medians/modes. The median CD4:CD8 ratio and P(CD4:CD8 $>1)$ were not
discernibly different between the three models.  The mean as a function of
age is only shown as derived from the uncensored Model 1.

\section{Discussion}

We have now established the asymptotic properties for censored CPMs, which
are flexible semiparametric regression models for continuous outcomes because
the outcome transformation is nonparametrically estimated.  We proved uniform
consistency of the estimated coefficients $\widehat\beta$ and the estimated
transformation function $\widehat A$ over the uncensored interval $[L,U]$,
and showed that their joint asymptotic distribution is a tight Gaussian
process.  We demonstrated that these estimators perform well with simulations
and illustrated their use in practice with a real data example.

Establishing uniform consistency requires a bounded range of the
transformation function $A$, which is achieved by censoring the outcome
variable at both ends.  Even if an outcome variable has a bounded support,
the transformed values may not be bounded, and censoring will still be needed
to establish uniform consistency.  The proof of uniform consistency for
$\widehat\beta$ also required a bounded range of $A$ even though $\beta$ and
$A$ are separate components of the model.

Although the asymptotic properties for a similar nonparametric maximum
likelihood approach in survival analysis have been established (Zeng and Lin,
2007), the proofs here for CPMs based on censored
data are different because we consider the nonparametric maximum likelihood
estimate for the transformation in CPMs rather than
the cumulative hazards function in survival analysis.  In addition, the
transformation is estimated in the proofs directionally and separately for
the two tails, which also differs from prior work.

For data without natural lower and upper bounds, the choice of $L$ and $U$
might be challenging in practice. In our CD4:CD8 ratio analysis, we were able
to select values of $L$ and $U$ that corresponded with small and large
CD4:CD8 percentiles in a prior study, therefore likely ensuring that a small
fraction of the data would be censored in our analysis. In general, it is
desirable to choose bounds so that only a small fraction of the data are
censored, although it should be reiterated that these bounds should be chosen
prior to analysis.  Both our simulations and data example suggest that
results are robust to the specific choices of $L$ and $U$ as long as they do
not severely censor data.  For example, in our simulations, results were
nearly identical when censoring varied between 0.2\% and 13\%; in the data
example, results were also nearly identical when censoring varied between
1.9\% and 11.4\%.

In addition, our simulations and data example actually suggest that without
censoring, the estimators also perform well, which may support the use of
uncensored CPMs in practice. Uncensored CPMs do not require specifying $L$
and $U$, and they permit calculation of conditional means. However, the
asymptotic theory presented here does not cover uncensored CPMs; hence, there
might be some risk to analyses using uncensored CPMs.

Continuous data that are skewed or subject to detection limits are common in
applied research. Because of their ability to non-parametrically estimate a
proper transformation, their robust rank-based nature, and their desirable
properties proved and illustrated in this manuscript, CPMs are often an
excellent choice for analyzing these types of data. Extensions of CPMs to
more complicated settings, e.g., clustered and longitudinal data,
multivariate outcome data, or data with multiple detection limits, are
warranted and are areas of ongoing research.

\section{Appendix}
\subsection*{A.1 Proof of Theorem 1}

Core steps of the proof: Let $(\beta_0, A_0)$ be the true value of
$(\beta, A)$, and $(\widehat\beta, \widehat A)$ be the NPMLE from the
censored approach of the CPM.  We will first prove that (I) $\widehat A$ is
bounded in $[L,U]$ with probability one.  Since $\widehat A(\cdot)$ is
bounded and increasing in $[L, U]$, by the Helly selection theorem, for any
subsequence, there exists a weakly convergent subsequence.  Thus, without
confusion, we assume that $\widehat A\rightarrow A^*$ weakly in $[L, U]$ and
$\widehat\beta \rightarrow\beta^*$.  We will then prove that (II) with
probability one, $A^*(y)=A_0(y)$ for $y\in [L, U]$ and $\beta^*=\beta_0$.
With this result, the consistency is established.  Furthermore, since $A_0$
is continuously differentiable, we conclude that $\widehat A(y)$ converges to
$A_0(y)$ uniformly in $[L, U]$ with probability one.

Proof of (I): Given a dataset of \text{i.i.d.} observations $\{(Y_i,Z_i)\}$,
the nonparametric log-likelihood for the censored approach of the CPM is
\begin{align}
  l_n(\beta, A) ={\bf P}_n \{
  &I(Y\le L)\log G(A(L)-\beta^TZ)
    +I(Y\ge U)\log (1-G(A(U-)-\beta^TZ)) \nonumber \\
  &\left.+I(L<Y<U)
    \log (G(A(Y)-\beta^TZ)-G(A(Y-)-\beta^TZ))\right\}.
\end{align}
Here, ${\bf P}_n$ denotes the empirical measure, i.e.,
${\bf P}_ng(Y,Z)=n^{-1}\sum_{i=1}^n g(Y_i, Z_i)$, for any measurable function
$g(Y,Z)$.  Let $\widehat A\{Y_i\}\equiv \widehat A(Y_i)-\widehat A(Y_i-)$ be
the jump size of $\widehat A$ at $Y_i$.  Let
$\widehat A(U)\equiv \widehat A(U-)$.

We first show that $\lim\sup \widehat A(U)<\infty$ a.s.  If
$\widehat A(Y_i-)\le M+m$ for all $Y_i>0$, then $\widehat A(U)\le M+m$.
Below we assume that is a $Y_i>0$ such that $\widehat A(Y_i-)>M+m$.  Clearly,
$\widehat A\{Y_i\}$ should be strictly positive, since otherwise,
$l_n(\beta, A)=-\infty$.  Because of this, we differentiate $l_n(\beta, A)$
with respect to $A\{Y_i\}$ and then set it to zero to obtain the following
equation:
$${\bf P}_n \left\{I(Y\ge U)\frac{G'(\widehat A(U)-\widehat\beta^TZ)}{1-G(\widehat A(U)-\widehat\beta^TZ)}\right\}-{\bf P}_n \left\{I(Y_i<Y<U)\frac{G'(\widehat A(Y)-\widehat\beta^TZ)-G'(\widehat A(Y-)-\widehat\beta^TZ)}{G(\widehat A(Y)-\widehat\beta^TZ)-G(\widehat A(Y-)-\widehat\beta^TZ)}\right\}$$
$$=\frac{1}{n}\frac{G'(\widehat A(Y_i)-\widehat\beta^TZ_i)}{G(\widehat A(Y_i)-\widehat\beta^TZ_i)-G(\widehat A(Y_i-)-\widehat\beta^TZ_i)}.\eqno(A.1)$$
For any $Y> Y_i$, by condition (C.2),
$$\widehat A(Y-)-\widehat\beta^TZ\ge M+m-\widehat\beta^TZ\ge M.$$
According to (C.1), $G'(x)$ is decreasing when $x\ge M$. The left-hand side
of (A.1) is
$$\ge{\bf P}_n \left\{I(Y>U)\frac{G'(\widehat
    A(U)-\widehat\beta^TZ)}{1-G(\widehat A(U)-\widehat\beta^TZ)}\right\}.$$ For the right-hand side, we use the mean-value theorem on the
denominator and then the decreasing property of $G'(x)$ when $x\ge M$ to
obtain
\begin{equation*}
  \frac{1}{n}\frac{G'(\widehat A(Y_i)-\widehat\beta^TZ_i)}{G(\widehat
    A(Y_i)-\widehat\beta^TZ_i)-G(\widehat A(Y_i-)-\widehat\beta^TZ_i)}
  =\frac{1}{n}\frac{G'(\widehat A(Y_i)-\widehat\beta^TZ_i)}{G'(\xi_i)\widehat A\{Y_i\}}
  \le\frac{1}{n\widehat A\{Y_i\}},
\end{equation*}
where $\xi_i$ is some value such that
$\widehat A(Y_i-)-\widehat\beta^TZ_i\le\xi_i\le\widehat
A(Y_i)-\widehat\beta^TZ_i$.  Therefore, we have
$$\widehat A\{Y_i\}\le \frac{1}{n}\left[{\bf P}_n \left\{I(Y>U)\frac{G'(\widehat A(U)-\widehat\beta^TZ)}{1-G(\widehat A(U)-\widehat\beta^TZ)}\right\}\right]^{-1},$$
and this holds for any $Y_i$ between $0$ and $U$ and satisfying
$\widehat A(Y_i-)>M+m$.

Let $i_0$ be the maximal index $i$ for which $Y_i>0$ and
$\widehat A(Y_{i_0}-)\le M+m$.  We sum over all $Y_i$ between 0 and $U$ to
obtain
\begin{eqnarray*}
  \widehat A(U)
  &=& \widehat A(Y_{i_0}-)+\sum_{Y_i>0, \widehat A(Y_i-)>M+m}\widehat A\{Y_i\}\\
  &\le& M+m+ \left[n^{-1}\sum_{i=1}^n I(0<Y_i\le U)\right]\left[{\bf P}_n \left\{I(Y>U)\frac{G'(\widehat A(U)-\widehat\beta^TZ)}{1-G(\widehat A(U)-\widehat\beta^TZ)}\right\}\right]^{-1}.
\end{eqnarray*}

We now show that $\widehat A(U)$ cannot diverge to $\infty$.  Otherwise,
suppose that $\widehat A(U)\rightarrow\infty$ for some subsequence.  From the
second half of Condition (C.1), when $n$ is large enough in the subsequence,
for any $Z$,
$$\frac{G'(\widehat A(U)-\widehat\beta^TZ)}{1-G(\widehat
  A(U)-\widehat\beta^TZ)}>\frac{1}{2}\lim\inf_{x\rightarrow\infty}\frac{G'(x)}{1-G(x)}
\equiv c_0>0,$$ and therefore,
$$\widehat A(U)\le M+m+ \frac{n^{-1}\sum_{i=1}^n I(0<Y_i\le U)}{c_0{\bf P}_n \left\{I(Y>U)\right\}},$$
in which the last term converges to a constant.  We thus have a
contradiction.  Hence, $\lim\sup \widehat A(U)<\infty$ with probability 1.

We can reverse the order of $Y_i$ (change $Y_i$ to $-Y_i$ so the NPMLE is
equivalent to maximizing the likelihood function but instead of $A(y)$, we
consider $-A(y)$).  The same arguments as above apply to conclude that
$\limsup -\widehat A(L)<\infty$ with probability 1, or equivalently,
$\liminf \widehat A(L)>-\infty$ with probability 1.

%Since $\widehat A(\cdot)$ is bounded and increasing in $[L, U]$, by the Helly selection theorem, for any subsequence, there exists a weakly convergent subsequence. Thus, without confusion, we assume that $\widehat A\rightarrow A^*$ weakly in $[L, U]$ and $\widehat\beta \rightarrow\beta^*$. The consistency is established once we can show $A^*(y)=A_0(y)$ for $y\in [L, U]$ and $\beta^*=\beta_0$, where $(\beta_0, A_0)$ is the true value for $(\beta, A).$

Proof of (II): We first show that $n\widehat A\{Y_i\}$ is bounded for all
$Y_i\in [L, U]$.  From the proof above, we know $\widehat A\{Y_i\}=O(n^{-1})$
uniformly in $i$ for which $Y_i\in [L, U]$ satisfying
$|\widehat A(Y_i-)|>M+m$. We prove that this is true for any $Y_i$.  To do
that, we define
\begin{eqnarray*}
H_n(y)&=&{\bf P}_n \left\{I(Y>U)\frac{G'(\widehat A(U)-\widehat \beta^TZ)}{1-G(\widehat A(U)-\widehat \beta^TZ)}\right\}\\
& &-{\bf P}_n \left\{I(y<Y\le U)\frac{G'( \widehat A(Y)-\widehat \beta^TZ)-G'(\widehat A(Y-)-\widehat \beta^TZ)}{G(\widehat A(Y)-\widehat \beta^TZ)-G(\widehat A(Y-)-\widehat \beta^TZ)}\right\}.
\end{eqnarray*}

First, we note that $H_n(y)$ has a total bounded variation in $[0,U]$. In fact,
for any $0<t<s<U$,
\begin{eqnarray*}
& &|H_n(t)-H_n(s)|\\
&=&\Big |{\bf P}_n \left\{I(t<Y\le s) \frac{G'( \widehat A(Y)-\widehat \beta^TZ)-G'(\widehat A(Y-)-\widehat\beta^TZ)}{G(\widehat A(Y)-\widehat\beta^TZ)-G(\widehat A(Y-)-\widehat \beta^TZ)}\right\}\Big |\\
&\le& c_1 {\bf P}_n \left\{I(t<Y\le s) \right\},
\end{eqnarray*}
where $c_1=\sup_{x\in [-m, c_0+m]}|G''(x)|/\inf_{x\in [-m, c_0+m]}|G'(x)|$. 
By choosing a subsequence, we assume that $H_n(y)$ converges weakly to $H^*(y)$. From the above inequality and taking limits, it is clear
$$|H^*(t)-H^*(s)|\le c_1 P(t<Y\le s)$$
so $H^*(y)$ is Lipschitz continuous in $y \in [0, U]$. The latter property ensures that $H_n(y)$ uniformly converges to $H^*(t)$ for $t\in [0, U]$.

According to equation (A.1), we know
$$|H_n(Y_i)|=\frac{1}{n}\frac{G'(\widehat A(Y_i)-\widehat\beta^TZ_i)}{G(\widehat A(Y_i)-\widehat\beta^TZ_i)-G(\widehat A(Y_i-)-\widehat\beta^TZ_i)}\ge \frac{c_2}{n\widehat A\{Y_i\}},\eqno(A.2)$$
where $c_2=\inf_{x\in [-m, c_0+m] G'(x)}/\sup_{x\in [-m, c_0+m]} G'(x)$. Thus,
$$\widehat A\{Y_i\}\ge \frac{c_2}{n} \frac{1}{|H_n(Y_i)|+\epsilon}$$
for any positive constant $\epsilon$. 
This gives 
$$\widehat A(U)\ge {c_2}{\bf P}_n\left[ \frac{I(Y\in [0, U])}{|H_n(Y)|+\epsilon}\right].\eqno(A.3)$$
Since $H_n(Y)$ has a bounded total variation, $\{|H_n(Y)|+\epsilon\}^{-1}$ belongs to a Glivenko--Cantelli class bounded by $1/\epsilon$ and it converges in $L_2(P)$-norm to $\{|H^*(Y)|+\epsilon\}^{-1}$. As a result, the right-hand side of (A.3) converges to $c_2E[I(Y\in [0, U])(|H^*(Y)|+\epsilon)^{-1}]$ so we obtain
$$c_0\ge c_2\int_0^U\frac{f_Y(y)}{|H^*(y)|+\epsilon}dy,$$
where $f_Y(y)$ is the marginal density of $Y$. Let $\epsilon$ decrease to zero then from the monotone convergence theorem, we conclude
$$\int_0^U\frac{f_Y(y)}{|H^*(y)|}dy\le \frac{c_0}{c_2}.\eqno(A.4)$$

We use (A.4) to show that $\min_{y\in [0,\tau]}|H^*(y)|>0$. Otherwise, since $H^*(y)$ is continuous, there exists some $y_0\in [0,\tau]$ such that
$H^*(y_0)=0$. However, since $H^*(y)$ is Lipschitz continuous at $y_0$, the left-hand side of (A.4) is at least 
larger than $\int_{y_0}^{y_0+\delta} \{c_1|y-y_0|\}^{-1}dy$ if $y_0<U$ or $\int_{y_0-\delta}^{y_0}\{c_1|y-y_0|\}^{-1}dy$ if $y_0>0$ for some small constant $\delta$. The latter integrals are infinity. We obtain the contradiction. Hence, we conclude that $H^*(y)$ is uniformly bounded away from zero when $y\in [0,U]$. Thus, when $n$ is large enough, $|\widehat H_n(Y_i)|$ is larger than a positive constant $c_3$ uniformly for all $Y_i>0$. 
From (A.1), we thus obtain 
$$c_3\le \frac{c_4}{n\widehat A\{Y_i\}},$$
where $c_4=\sup_{x\in [-m, c_0+m] G'(x)}/\inf_{x\in [-m, c_0+m]}G'(x)$. In other words, 
$n\widehat A\{Y_i\}\le c_4/c_3$.
By symmetric arguments,  we can show that $n\widehat A\{Y_i\}$ is bounded by a constant for all $Y_i<0$

Finally, to establish the consistency in Theorem 1, since $\widehat A\{Y_i\}$ is of order $n^{-1}$, from equation (A.1), we obtain
$${\widehat A\{Y_i\}}=n^{-1}\left[{\bf P}_n \left\{I(Y>U)\frac{G'(\widehat A(U)-\widehat\beta^TZ)}{1-G(\widehat A(U)-\widehat\beta^TZ)}\right\}-{\bf P}_n \left\{I(Y_i<Y\le U)\frac{G''(\widehat A(Y)-\widehat\beta^TZ)}{G'(\widehat A(Y)-\widehat\beta^TZ)}\right\}\right]^{-1}$$
$$+O(n^{-2}).$$
Following this expression, we define another step function, denoted by $\widetilde A(y)$, whose jump size at $Y_i$ satisfies
$${\widetilde A\{Y_i\}}=n^{-1}\left[{\bf P}_n \left\{I(Y>U)\frac{G'(A_0(U)-\beta_0^TZ)}{1-G( A_0(U)-\beta_0^TZ)}\right\}-{\bf P}_n \left\{I(Y_i<Y\le U)\frac{G''( A_0(Y)-\beta_0^TZ)}{G'( A_0(Y)-\beta_0^TZ)}\right\}\right]^{-1}$$
so 
$$\widetilde A(y)=n^{-1}\sum_{i=1}^n I(Y_i\le y)\left[{\bf P}_n \left\{I(Y>U)\frac{G'(A_0(U)-\beta_0^TZ)}{1-G( A_0(U)-\beta_0^TZ)}\right\}\right.$$
$$\left.-{\bf P}_n \left\{I(Y_i<Y\le U)\frac{G''( A_0(Y)-\beta_0^TZ)}{G'( A_0(Y)-\beta_0^TZ)}\right\}\right]^{-1}.$$
By the strong law of large numbers and monotonicity of $\widehat A$, it is straightforward to show $\widehat A(y)$ converges to 
$$E\left\{ I(Y\le y)\left[\widetilde {\bf P} \left\{I(\widetilde Y>U)\frac{G'(A_0(U)-\beta_0^T\widetilde Z)}{1-G( A_0(U)-\beta_0^T\widetilde Z)}\right\}-\widetilde {\bf P} \left\{I(Y<\widetilde Y\le U)\frac{G''( A_0(\widetilde Y)-\beta_0^T\widetilde Z)}{G'( A_0(\widetilde Y)-\beta_0^T\widetilde Z)}\right\}\right]^{-1}\right\}$$
 uniformly in $y\in [L, U]$. The limit can be verified to be the same as $A_0(y)$. Furthermore, we notice
$$\frac{\widehat A\{Y_i\}}{\widetilde A\{Y_i\}}
=\frac{{\bf P}_n \left\{I(Y>U)\frac{G'(A_0(U)-\beta_0^TZ)}{1-G( A_0(U)-\beta_0^TZ)}\right\}-{\bf P}_n \left\{I(Y_i<Y\le U)\frac{G''( A_0(Y)-\beta_0^TZ)}{G'( A_0(Y)-\beta_0^TZ)}\right\}}{
\left[{\bf P}_n \left\{I(Y>U)\frac{G'(\widehat A(U)-\widehat\beta^TZ)}{1-G(\widehat A(U)-\widehat\beta^TZ)}\right\}-{\bf P}_n \left\{I(Y_i<Y\le U)\frac{G''(\widehat A(Y)-\widehat\beta^TZ)}{G'(\widehat A(Y)-\widehat\beta^TZ)}\right\}\right]+O(n^{-1})}.\eqno(A.5)$$

Since $\widehat A(y)$ is bounded and increasing and $\widehat\beta^TZ$, $\widehat A(Y)-\widehat \beta^TZ$ belongs to a VC-hull so Donsker class. By the preservation property under the monotone transformation,   $G^{(k)}(\widehat A(U)-\widehat\beta^TZ)$, $k=0,1,2$, also belongs to a Donsker class. Therefore, the right-hand side of (A.5) converges uniformly in $Y_i$ to
$$g(Y_i)=\frac{{\bf P} \left\{I(Y>U)\frac{G'(A_0(U)-\beta_0^TZ)}{1-G( A_0(U)-\beta_0^TZ)}\right\}-{\bf P} \left\{I(Y_i<Y\le U)\frac{G''( A_0(Y)-\beta_0^TZ)}{G'( A_0(Y)-\beta_0^TZ)}\right\}}{
{\bf P} \left\{I(Y>U)\frac{G'( A^*(U)-{\beta^*}^TZ)}{1-G(A^*(U)-{\beta^*}^TZ)}\right\}-{\bf P} \left\{I(Y_i<Y\le U)\frac{G''(A^*(Y)-{\beta^*}^TZ)}{G'(A^*(Y)-{\beta^*}^TZ)}\right\}}.$$
As a result, $A^*(y)=\int_0^y g(t)dA_0(t)$, or equivalently, $dA^*(y)/dA_0(y)=g(y)$.

Define
$$\widetilde l_n(\beta, A)={\bf P}_n\left\{I(Y\le L)\log G(A(L)-\beta^TZ)+I(Y>U)\log (1-G(A(U)-\beta^TZ))\right.$$
$$\left.+I(L<Y\le U)
\log (G(A(Y)-\beta^TZ)A\{Y\}\right\}.$$
Since $\widetilde A\{Y_i\}=O(n^{-1})$ and $\widehat A\{Y_i\}=O(n^{-1})$,
$$l_n(\widehat\beta, \widehat A)=\widetilde l_n(\widehat\beta, \widehat A)+O(n^{-1}), \ \
l_n(\beta_0, \widetilde A)=\widetilde l_n(\beta_0, \widetilde A)+O(n^{-1}).$$
Since  $l_n(\widehat\beta, \widehat A)\ge l_n(\beta_0, \widetilde A_0),$ we have
$$\widetilde l_n(\widehat\beta, \widehat A)\ge \widetilde l_n(\beta_0, \widetilde A)+O(n^{-1}).$$
That is,
$${\bf P}_n\left\{I(Y\le L)\log \frac{G(\widehat A(L)-\widehat \beta^TZ)}{G(\widetilde A(L)-\beta_0^TZ)}+I(Y>U)\log \frac{1-G(\widehat A(U)-\widehat\beta^TZ)}{1-G(\widetilde A(U)-\beta_0^TZ)}\right\}$$
$$+{\bf P}_n \left\{I(L<Y\le U)
\log \frac{G(\widehat A(Y)-\widehat \beta^TZ)}{G(\widetilde A(Y)-\beta_0^TZ)}\right\}+n^{-1}\sum_{i=1}^n I(L<Y_i\le U)\frac{\widehat A\{Y_i\}}{\widetilde A\{Y_i\}}\ge O(n^{-1}).$$
We take limits on both sides. Using the Glivenko--Cantelli theorem to the first three two terms in the left-hand side and noting $\Big |\widehat A\{Y_i\}/\widetilde A\{Y_i\}-g(Y_i)\Big |$ converges to zero uniformly, we obtain
$${\bf P}\left\{I(Y\le L)\log \frac{G(A^*(L)-{\beta^*}^TZ)}{G(A_0(L)-\beta_0^TZ)}+I(Y>U)\log \frac{1-G(A^*(U)-{\beta^*}^TZ)}{1-G(A_0(U)-\beta_0^TZ)}\right\}$$
$$+{\bf P} \left\{I(L<Y\le U)
\log \frac{G(A^*(Y)-{\beta^*}^TZ)}{G(A_0(Y)-\beta_0^TZ)}\right\}+{\bf P}\left\{(L<Y\le U)\frac{d A^*(Y)}{dA_0(Y)}\right\}\ge 0.$$
The left-hand side is the negative Kullback--Leibler information for the density with parameter $(\beta^*, A^*)$. Thus, the density function with parameter $(\beta^*, A^*)$ should be the same as the true density. Immediately, we obtain
$$G(A^*(Y)+{\beta^*}^TZ)=G(A_0(Y)+\beta_0^TZ)$$
with probability one. From condition C.2, we conclude that $\beta^*=\beta_0$ and $A^*(y)=A_0(y)$ for $y\in [L, U]$. 
%The consistency holds. Furthermore, since $A_0$ is continuously differentiable, we also conclude that $\widehat A(y)$ converges to $A_0(y)$ uniformly in $[L, U]$ with probability one.

\subsection*{A.2 Proof of Theorem 2}
\allowdisplaybreaks

Let $BV[L,U]$ be the set of the functions over $[L,U]$ with
$\Vert h\Vert_{TV}\le 1$, where ${\Vert\cdot\Vert_{TV}}$ denotes the total
variation in $[L,U]$.  For any $\nu\in R^p$ with $\Vert \nu\Vert\le 1$ and
any $h\in BV[L,U]$, we define the score function $\bPsi_n(\beta, A)[\nu, h]$
along the submodel for $\beta$ with tangent direction $\nu$ and for $A$ with
the tangent function $\int_0^{\cdot} h(t)dA(t)$:
\begin{eqnarray*}
  & &\bPsi_n(\beta, A)[\nu, h]\\
  &=&\lim_{\epsilon\rightarrow 0}
      \left(\frac{l_n(\beta+\epsilon \nu,  A+\epsilon \int_0^{\cdot} h(t)dA(t))-l_n(\beta, A)}{\epsilon}\right)\\
  &=& {\bf P}_n\left\{ -F_1(Y,Z;\beta, A)Z^T\nu
      +F_2(Y,Z;\beta, A)Z^T\nu
      -F_3(Y,Z;\beta, A)Z^T\nu\right\} \\
  & & + {\bf P}_n\left\{ F_1(Y,Z;\beta, A)\int_0^L h(t)dA(t)
      -F_2(Y,Z;\beta, A)\int_0^Uh(t)dA(t)\right.\\
  & & \qquad\qquad \left.
      +F_3(Y,Z;\beta, A)\int_0^YhdA+F_4(Y,Z; \beta, A)h(Y)\right\},
\end{eqnarray*}
where
\begin{eqnarray*}
  F_1(Y, Z; \beta, A)&=&\frac{I(Y\le L)G'(A(L)-\beta^TZ)}{G(A(L)-\beta^TZ)},\\
  F_2(Y, Z; \beta, A)&=&\frac{I(Y\ge U)G'(A(U)-\beta^TZ)}{1-G(A(U)-\beta^TZ)},\\
  F_3(Y,Z;\beta, A)&=&\frac{I(L<Y<U)\left(G'(A(Y)-\beta^TZ)-G'(A(Y-)-\beta^TZ)\right)}{G(A(Y)-\beta^TZ)-G(A(Y-)-\beta^TZ)},\\
  F_4(Y,Z;\beta, A)&=&\frac{G'(A(Y-)-\beta^TZ)}{G(A(Y)-\beta^TZ)-G(A(Y-)-\beta^TZ)}(A(Y)-A(Y-)).
\end{eqnarray*}
Since $(\widehat\beta, \widehat A)$ maximizes $l_n(\beta, A)$, we have, for
any $v$ and $h$,
$$\bPsi_n(\widehat \beta, \widehat A)[\nu, h]=0.$$

The rest of the proof contains the following main steps: we first show that
$(\widehat\beta, \widehat A)$ satisfies equation (A.6) (details below), and
then (A.8) and finally (A.10), from which the asymptotic distribution of
$(\widehat\beta, \widehat A)$ will be derived.

We know $\max_{L\le Y_i\le U}(\widehat A(Y_i)-\widehat A(Y_i-))=O_p(n^{-1})$
from the proof in Section A.1.  Thus, if we let
$$\widetilde F_3(Y, Z; \beta, A)=\frac{I(L<Y<U)G''(A(Y)-\beta^TZ)}{G'(A(Y)-\beta^TZ)},$$
then
$$F_3(Y,Z; \widehat\beta, \widehat A)=\widetilde F_3(Y, Z;\widehat\beta, \widehat A)+O_p(n^{-1})$$
and $F_4(Y, Z;\widehat\beta, \widehat A)=1+O_p(n^{-1})$ uniformly in $(Y, Z)$.
Consequently,
we obtain
\begin{eqnarray*}
& &{\bf P}_n\left(
 -F_1(Y,Z;\widehat\beta, \widehat A)Z^T\nu
+F_2(Y,Z;\widehat\beta, \widehat A)Z^T\nu
-\widetilde F_3(Y,Z;\widehat\beta, \widehat A)Z^T\nu\right)
\\
& &+{\bf P}_n\left(F_1(Y,Z;\widehat\beta, \widehat A)\int_0^L h(t)d\widehat A(t)
-F_2(Y,Z;\widehat\beta, \widehat A)\int_0^Uh(t)d\widehat A(t)\right.\\
& &\qquad \qquad \left.
\qquad\qquad 
+\widetilde F_3(Y,Z;\widehat\beta, \widehat A)\int_0^Yhd\widehat A+h(Y)
\right)\\
&=&O_p(n^{-1}),
\end{eqnarray*}
and it holds uniformly in $\nu$ and $h$ with $\Vert\nu\Vert\le 1$ and $\Vert h\Vert_{TV}\le 1$.
Equivalently,
we have
$$\sqrt n({\bf P}_n-{\bf P})\left(
\begin{matrix}
 -F_1(Y,Z;\widehat\beta, \widehat A)Z^T\nu
+F_2(Y,Z;\widehat\beta, \widehat A)Z^T\nu
-\widetilde F_3(Y,Z;\widehat\beta, \widehat A)Z^T\nu
\cr
+ F_1(Y,Z;\widehat\beta, \widehat A)\int_0^L h(t)d\widehat A(t)
-F_2(Y,Z;\widehat\beta, \widehat A)\int_0^Uh(t)d\widehat A(t)\cr
\qquad\qquad 
+\widetilde F_3(Y,Z;\widehat\beta, \widehat A)\int_0^Yhd\widehat A+h(Y)
\end{matrix}
\right)
$$
$$=\sqrt n {\bf P}\left(
\begin{matrix}
 -F_1(Y,Z;\widehat\beta, \widehat A)Z^T\nu
+F_2(Y,Z;\widehat\beta, \widehat A)Z^T\nu
-\widetilde F_3(Y,Z;\widehat\beta, \widehat A)Z^T\nu
\cr
 +F_1(Y,Z;\widehat\beta, \widehat A)\int_0^L h(t)d\widehat A(t)
-F_2(Y,Z;\widehat\beta, \widehat A)\int_0^Uh(t)d\widehat A(t)\cr
\qquad\qquad 
+\widetilde F_3(Y,Z;\widehat\beta, \widehat A)\int_0^Yhd\widehat A+h(Y)
\end{matrix}
\right)+O_p(n^{-1/2}).\eqno(A.6)$$

For the left-hand side of (A.6), it is easy to see that
for $(\beta, A)$ in a neighborhood of $(\beta_0, A_0)$ in the metric space $R^d\times BV[L, U]$,
the classes of $F_1(Y,Z;\beta, A)$, $F_2(Y,Z;\beta, A)$ and $\widetilde F_3(Y,Z;\beta, A)$ are Lipschtiz classes of the P-Donsker classes $\left\{\beta^TZ\right\}$ and 
$\left\{A(Y)\in BV[L, U]\right\}$, so they are P-Donsker by preservation of the Donsker property. Additionally, the classes of $\left\{\int_0^YhdA\right\}$, $\left\{Z^T\nu\right\}$ and $\left\{h(Y)\right\}$ are P-Donsker.
As the result, 
since by the consistency,
$$\left(
\begin{matrix}
 -F_1(Y,Z;\widehat\beta, \widehat A)Z^T\nu
+F_2(Y,Z;\widehat\beta, \widehat A)Z^T\nu
-\widetilde F_3(Y,Z;\widehat\beta, \widehat A)Z^T\nu
\cr
 +F_1(Y,Z;\widehat\beta, \widehat A)\int_0^L h(t)d\widehat A(t)
-F_2(Y,Z;\widehat\beta, \widehat A)\int_0^Uh(t)d\widehat A(t)\cr
\qquad\qquad 
+\widetilde F_3(Y,Z;\widehat\beta, \widehat A)\int_0^Yhd\widehat A+h(Y)
\end{matrix}
\right)
$$
converges in $L_2(P)$ to
$$\bS(Y,Z)[\nu, h]\equiv\left(
\begin{matrix}
 -F_1(Y,Z;\beta_0, A_0)Z^T\nu
+F_2(Y,Z;\beta_0, A_0)Z^T\nu
-\widetilde F_3(Y,Z;\beta_0, A_0)Z^T\nu
\cr
+ F_1(Y,Z;\beta_0, A_0)\int_0^L h(t)dA_0(t)
-F_2(Y,Z;\beta_0, A_0)\int_0^Uh(t)dA_0(t)\cr
\qquad\qquad 
+\widetilde F_3(Y,Z;\beta_0, A_0)\int_0^YhdA_0+h(Y)
\end{matrix}
\right),$$
equation (A.6) gives
$$\sqrt n ({\bf P}_n-{\bf P})\bS(Y,Z)[\nu, h]$$
$$=\sqrt n {\bf P}\left(
\begin{matrix}
 -F_1(Y,Z;\widehat\beta, \widehat A)Z^T\nu
+F_2(Y,Z;\widehat\beta, \widehat A)Z^T\nu
-\widetilde F_3(Y,Z;\widehat\beta, \widehat A)Z^T\nu
\cr
 F_1(Y,Z;\widehat\beta, \widehat A)\int_0^L h(t)d\widehat A(t)
-F_2(Y,Z;\widehat\beta, \widehat A)\int_0^Uh(t)d\widehat A(t)\cr
\qquad\qquad 
+\widetilde F_3(Y,Z;\widehat\beta, \widehat A)\int_0^Yhd\widehat A+h(Y)
\end{matrix}
\right)+o_p(1).\eqno(A.7)$$

On the other hand, we note that the first term in the right-hand side of (A.7) is zero if replacing $(\widehat\beta, \widehat A)$ by $(\beta_0, A_0)$. Thus, the right-hand side of (A.7) is equal to
$$
\sqrt n {\bf P}\left(
\begin{matrix}
 -F_1(Y,Z;\widehat\beta, \widehat A)Z^T\nu
+F_2(Y,Z;\widehat\beta, \widehat A)Z^T\nu
-\widetilde F_3(Y,Z;\widehat\beta, \widehat A)Z^T\nu
\cr
 +F_1(Y,Z;\widehat\beta, \widehat A)\int_0^L h(t)d\widehat A(t)
-F_2(Y,Z;\widehat\beta, \widehat A)\int_0^Uh(t)d\widehat A(t)\cr
\qquad\qquad 
+\widetilde F_3(Y,Z;\widehat\beta, \widehat A)\int_0^Yhd\widehat A
\end{matrix}
\right)$$
$$
-\sqrt n {\bf P}\left(
\begin{matrix}
 -F_1(Y,Z;\beta_0, A_0)Z^T\nu
+F_2(Y,Z;\beta_0, A_0)Z^T\nu
-\widetilde F_3(Y,Z;\beta_0, A_0)Z^T\nu
\cr
+ F_1(Y,Z;\beta_0, A_0)\int_0^L h(t)dA_0(t)
-F_2(Y,Z;\beta_0, A_0)\int_0^Uh(t)dA_0(t)\cr
\qquad\qquad 
+\widetilde F_3(Y,Z;\beta_0, A_0)\int_0^YhdA_0
\end{matrix}
\right)+o_p(1).$$
We perform the linearization to the first two terms in the above expression. After some algebra, we obtain that this expression is equivalent to
$$\sqrt n (\bS_{11}^T\nu+\bS_{12}[h])^T(\widehat\beta-\beta_0)
+\sqrt n\int (\bS_{21}^T\nu+\bS_{22}[h]) d(\widehat A-A_0)(y)$$
$$+ o_p(\sqrt n \Vert \widehat\beta-\beta_0\Vert+\sqrt n \Vert \widehat
A-A_0\Vert_{TV})+o_p(1),$$ where the operators
$\bS_{11}: R^d\rightarrow R^d$, $\bS_{12}: BV[L,U]\rightarrow R^d$,
$\bS_{21}^T: R^d\rightarrow BV[L,U]$, and
$\bS_{22}: BV[L,U]\rightarrow BV[L,U]$ are defined as
\begin{eqnarray*}
  \bS_{11}v
  &=&E\left[\frac{d}{dt}\frac{G'(t)}{G(t)}\Big |_{t=A_0(L)-\beta_0^TZ}I(Y\le L)ZZ^Tv\right]\\
  & &-E\left[\frac{d}{dt}\frac{G'(t)}{1-G(t)}\Big |_{t=A_0(U)-\beta_0^TZ}I(Y\ge U)ZZ^Tv\right]\\
  & &+E\left[\frac{d}{dt}\frac{G''(t)}{G'(t)}\Big |_{t=A_0(Y)-\beta_0^TZ}I(L<Y<U)ZZ^Tv\right],\\
  \bS_{12}[h]
  &=&-E\left[\frac{d}{dt}\frac{G'(t)}{G(t)}\Big |_{t=A_0(L)-\beta_0^TZ}I(Y\le L)Z\int_0^L hdA_0\right]\\
  & &+E\left[\frac{d}{dt}\frac{G'(t)}{1-G(t)}\Big |_{t=A_0(U)-\beta_0^TZ}I(Y\ge U)Z\int_0^UhdA_0\right]\\
  & &-E\left[\frac{d}{dt}\frac{G''(t)}{G'(t)}\Big |_{t=A_0(Y)-\beta_0^TZ}I(L<Y<U)Z\int_0^YhdA_0\right],\\
  (\bS_{21}^Tv)(y)
  &=&-E\left[\frac{d}{dt}\frac{G'(t)}{G(t)}\Big |_{t=A_0(L)-\beta_0^TZ}I(Y\le L)Z^TvI(Y>y)\right]\\
  & &+E\left[\frac{d}{dt}\frac{G'(t)}{1-G(t)}\Big |_{t=A_0(U)-\beta_0^TZ}I(Y\ge U)Z^TvI(Y>y)\right]\\
  & &-E\left[\frac{d}{dt}\frac{G''(t)}{G'(t)}\Big |_{t=A_0(Y)-\beta_0^TZ}I(L<Y<U)Z^TvI(Y>y)\right], \\
  \bS_{22}[h](y)
  &=&E\left[\frac{d}{dt}\frac{G'(t)}{G(t)}\Big |_{t=A_0(L)-\beta_0^TZ}I(Y\le L)I(Y>y)\int_0^L hdA_0\right]\\
  & &-E\left[\frac{d}{dt}\frac{G'(t)}{1-G(t)}\Big |_{t=A_0(U)-\beta_0^TZ}I(Y\ge U)I(Y>y)\int_0^UhdA_0\right]\\
  & &+E\left[\frac{d}{dt}\frac{G''(t)}{G'(t)}\Big |_{t=A_0(Y)-\beta_0^TZ}I(L<Y<U)I(Y>y)\int_0^YhdA_0\right]\\
  & &+ E\left\{F_1(Y,Z;\beta_0,A_0)I(L\ge y)-F_2(Y,Z;\beta_0,A_0)I(U>y)+\widetilde F_3(Y,Z;\beta_0,A_0)I(Y\le y)\right\}h(y).
\end{eqnarray*}
Combining the above results, we obtain from (A.7) that
$$\sqrt n (\bS_{11}^T\nu+\bS_{12}[h])^T(\widehat\beta-\beta_0)
+\sqrt n\int (\bS_{21}^T\nu+\bS_{22}[h]) d(\widehat A-A_0)(y)$$
$$=\sqrt n ({\bf P}_n-{\bf P})\bS(Y,Z)[\nu, h]+ o_p(\sqrt n \Vert \widehat\beta-\beta_0\Vert+\sqrt n \Vert \widehat
A-A_0\Vert_{TV})+o_p(1).\eqno(A.8)$$

Next, we show the operator,
$(\bS_{11}^T\nu+\bS_{12}[h], \bS_{21}^T\nu+\bS_{22}[h])$ that maps
$(\nu, h)\in R^d\times BV[L, U]$ to $R^d\times BV[L, U]$, is invertible.
This can be proved as follows: first, $\bS_{11}^T\nu+\bS_{12}[h]$ is finite
dimensional. Second, since the last term of $\bS_{22}[h]$ is invertible and
the other terms in $\bS_{21}^T\nu+\bS_{22}[h]$ maps $(\nu, h)$ to a
continuously differentiable function in $[L, U]$ which is a compact operator,
$\bS_{21}^T\nu+\bS_{22}[h]$ is a Fredholm operator of the first
kind. Therefore, to prove the invertibility, it suffices to show that
$(\bS_{11}^T\nu+\bS_{12}[h], \bS_{21}^T\nu+\bS_{22}[h])$ is one to
one. Suppose that $(\bS_{11}^T\nu+\bS_{12}[h],
\bS_{21}^T\nu+\bS_{22}[h])=0$. Thus, we have
$$(\bS_{11}^T\nu+\bS_{12}[h])^T\nu+\int (\bS_{21}^T\nu+\bS_{22}[h]) dh(y)=0.$$
From the previous derivation, we note that the left-hand side is in fact the negative Fisher information along the submodel $(\beta_0+\epsilon \nu, 
A_0+\int_0^{\cdot}h(t)dA_0(t))$. Thus, the score function along this submodel must be zero almost surely. That is,
$$-F_1(Y,Z;\beta_0, A_0)Z^T\nu
+F_2(Y,Z;\beta_0, A_0)Z^T\nu
-F_3(Y,Z;\beta_0, A_0)Z^T\nu
$$
$$
+F_1(Y,Z;\beta_0, A_0)\int_0^L h(t)dA(t)
-F_2(Y,Z;\beta_0, A_0)\int_0^Uh(t)dA(t)
+\widetilde F_3(Y,Z;\beta_0, A_0)\int_0^YhdA+h(Y)=0$$
almost surely. Consider any $Y=0$ then we have $Z^T\nu-h(0)=0$ so $\nu=0$ from Condition C.2. This further gives that $h(0)=0$ and $h(y)$ satisfies
$$h(Y)+\widetilde F_3(Y,Z;\beta_0, A_0)\int_0^YhdA=0$$
for any $Y\in [L, U]$. This is a homogeneous integral equation and it is clear $h(y)=0$ for any $y\in [L, U]$. We thus have established the invertibility of the operator $(\bS_{11}^T\nu+\bS_{12}[h], \bS_{21}^T\nu+\bS_{22}[h])$.

Therefore, from (A.8), for any $\nu^*$ and $h^*$, by choosing $(\nu^-, h^-)$ as the inverse of the above operator applied to $(\nu^*, h^*)$, we obtain
$$\sqrt n {\nu^*}^T(\widehat\beta-\beta_0)
+\sqrt n\int h^*(y) d(\widehat A-A_0)(y)$$
$$=\sqrt n ({\bf P}_n-{\bf P})\bS(Y,Z)[\nu^-, h^-]+ o_p(\sqrt n \Vert \widehat\beta-\beta_0\Vert+\sqrt n \Vert \widehat
A-A_0\Vert_{TV})+o_p(1)\eqno(A.9)$$ and this holds uniformly for
$\Vert\nu^*\Vert\le 1$ and $\Vert h^*\Vert_{TV}\le 1$.  Using (A.9), we
obtain
$$\sqrt n \Vert \widehat\beta-\beta_0\Vert+\sqrt n \Vert \widehat
A-A_0\Vert_{TV}=O_p(1).$$
Thus,
$$\sqrt n {\nu^*}^T(\widehat\beta-\beta_0)
+\sqrt n\int h^*(y) d(\widehat A-A_0)(y)$$
$$=\sqrt n ({\bf P}_n-{\bf P})\bS(Y,Z)[\nu^-, h^-]+o_p(1).\eqno(A.10)$$
This implies that
$$\sqrt n (\widehat\beta-\beta_0, \widehat A-A_0),$$
as a random map on $(\nu^*, h^*)$,
converges weakly to a mean-zero and tight Gaussian process. Furthermore, by letting $\nu^*=1$ and $h^*=0$, we conclude that $\sqrt n (\widehat\beta-\beta_0)$ has an influence function given by $\bS(Y,Z)[\nu^-, h^-]$. Since the latter lies on the score space, it must be the efficient influence function. Hence, the asymptotic variance of $\sqrt n (\widehat\beta-\beta_0)$ achieve the semiparametric efficiency bound.

\section*{Acknowledgements}

Jessica Castilho and other Vanderbilt Comprehensive Care Clinic investigators
for use for CD4:CD8 data.  This study was supported in part by United States
National Institutes of Health grants R01AI093234, P30AI110527, and
K23AI20875.

Detailed proofs of Theorems 1 and 2.  Additional results from simulations and
data example.  The code for simulations and data analysis
is available at \\
\texttt{https://biostat.app.vumc.org/ArchivedAnalyses}.

\section*{References}

Cheng SC, Wei LJ, Ying Z.  Analysis of transformation models with censored
data. {\it Biometrika}. 1995;82(4):835--845.

Hothorn T, M\"ost L, B\"uhlmann P. Most likely transformations. {\it
  Scandinavian Journal of Statistics}. 2017;45:110--134.

Cox DR. Regression models and life-tables. {\it J R Stat Soc Series B
  (Methodol)}. 1972;34(2):187--220.

Zeng D, Lin DY. Maximum likelihood estimation in semiparametric regression
models with censored data. {\it J R Stat Soc Series B (Methodol)}.
2007;69(4):507--564.

Liu Q, Shepherd BE, Li C, Harrell Jr FE. Modeling continuous response
variables using ordinal regression. {\it Statistics in
  Medicine}. 2017;36:4316--4335.

van der Vaart AW, Wellner JA. {\it Weak Convergence and Empirical
  Processes}. 1996. Springer

Murphy SA, van der Vaart AW (2000) On Profile Likelihood. {\it Journal of the
  American Statistical Association}. 2000;95(450):449--465.

Zeng D, Lin DY. Efficient estimation of semiparametric transformation models
for counting processes. {\it Biometrika}. 2006:93(3):627--640.

Castilho JL, Shepherd BE, Koethe J, et al. (2016) CD4/CD8 ratio, age, and risk of serious non-communicable diseases in HIV-infected adults on antiretroviral therapy. \emph{AIDS} 30: 899--908.

Sauter R, Huang R, Ledergerber B, et al. (2016). CD4/CD8 ratio and CD8 counts predict CD4 response in HIV-1-infected drug naive and in patients on cART. \emph{Medicine} 95: e5094.

Petoumenos K, Choi JY, Hoy J, et al. (2017). CD4:CD8 ratio comparison between cohorts of HIV-positive Asians and Caucasians upon commencement of antiretroviral therapy. \emph{Antiviral Therapy} 22: 659--668.

Serrano-Villar, Sainz T, Lee SA, et al. (2015) HIV-infected individuals with low CD4/CD8 ratio despite effective antiretroviral therapy exhibit altered T cell subsets, heightened CD8+ T cell activation, and increased risk of non-AIDS morbidity and mortality. \emph{PLOS Pathogens} 10: e1004078.

Silva C, Peder L, Silva E, Previdelli I, Pereira O, Teixeira J, Bertolini D (2018). Impact of HBV and HCV coinfection on CD4 cells among HIV-infected patients: a longitudinal retrospective study. \emph{The Journal of Infection in Developing Countries} 12: 1009--1018.

Gras L, May M, Ryder LP, et al. (2019). Determinants of restoration of CD4 and CD8 cell counts and their ratio in HIV-1-positive individuals with sustained virological suppression on antiretroviral therapy. \emph{Journal of Acquired Immune Deficiency Syndrome} 80: 292--300.

Serrano-Villar, Perez-Elias MJ, Dronda F, et al. (2014). Increased risk of serious non-AIDS-related events in HIV-infected subjects on antiretroviral therapy associated with a low CD4/CD8 ratio. \emph{PLOS ONE}. 9: e85798.

Shephard BE, Li C, Liu Q. Probability-scale residuals. \emph{Canadian Journal
  of Statistics}.  2016;44(4):463--479.


\begin{thebibliography}{7}
\expandafter\ifx\csname natexlab\endcsname\relax\def\natexlab#1{#1}\fi

\bibitem[{Castilho(2016)}]{Castilho:2016}
\textsc{Castilho, J. L., Shepherd, B. E., Koethe, J. R., Turner, M., Bebawy, S., Logan, J., Rogers, W. B., Raffanti, S. \& Sterling, T. R.} (2016).
\newblock {CD4/CD8 ratio, age, and risk of serious non-communicable diseases in HIV-infected adults on antiretroviral therapy}.
\newblock \textit{AIDS} \textbf{30}, 899--908.


\bibitem[{Cheng(1995)}]{cheng:1995}
\textsc{Cheng, S. C., Wei, L. J. \& Ying, Z.} (1995).
\newblock {Analysis of transformation models with censored data.} 
\newblock \textit{Biometrika} \textbf{82}, 835--845.

\bibitem[{Cox(1972)}]{Cox:1972}
\textsc{Cox, D.~R.} (1972).
\newblock {Regression models and life tables (with Discussion)}.
\newblock \textit{J. R. Statist. Soc. {\rm B}} \textbf{34}, 187--220.

\bibitem[{Gras(2019)}]{gras:2019}
\textsc{Gras, L., May, M., Ryder, L. P., Trickey, A., Helleberg, M., Obel, N., Thiebaut, R., Guest, J., Gill, J., Crane, H., Dias Lima, V., d'Arminio Monforte, A., Sterling, T. R., Miro, J., Moreno, S., Stephan, C., Smith, C., Tate, J., Shepherd, L., Saag, M., Rieger, A., Gillor, D., Cavassini, M., Montero, M., Ingle, S. M., Reiss, P., Costagliola, D., Wit, F. W. N. M., Sterne, J., de Wolf, F. \& Geskus, R.} (2019). 
\newblock {Determinants of restoration of CD4 and CD8 cell counts and their ratio in HIV-1-positive individuals with sustained virological suppression on antiretroviral therapy}. 
\newblock \textit{J. Acquir. Immune Defic. Syndr.} \textbf{80}, 292--300.

\bibitem[{Hothorn(2017)}]{hothorn:2017}
\textsc{Hothorn, T., M\"ost, L. \& B\"uhlmann, P.} (2017). 
\newblock {Most likely transformations.} 
\newblock \textit{Scand. J. Stat.} \textbf{45}, 110--134.

\bibitem[{Liu(2017)}]{Liu:2017}
\textsc{Liu, Q., Shepherd, B. E., Li, C. \& Harrell, F. E.} (2017).
\newblock {Modeling continuous response variables using ordinal regression}.
\newblock \textit{Stat. Med.} \textbf{36}, 4316--4335.

\bibitem[{Murphy(2000)}]{Murphy:2000}
\textsc{Murphy, S. A. \& van der Vaart, A. W.} (2000).
\newblock {On profile likelihood}.
\newblock \textit{J. Am. Stat. Assoc.} \textbf{95}, 449--465.

\bibitem[{Petoumenos(2017)}]{Petoumenos:2017}
\textsc{Petoumenos, K., Choi, J. Y., Hoy, J., Kiertiburanakul, S., Ng, O. T., Boyd, M., Rajasuriar, R. \& Law, M.} (2017).
\newblock {CD4:CD8 ratio comparison between cohorts of HIV-positive Asians and Caucasians upon commencement of antiretroviral therapy}.
\newblock \textit{Antiviral Therapy} \textbf{22}, 659--668.


\bibitem[{Sauter(2016)}]{Sauter:2016}
\textsc{Sauter, R., Huang, R., Ledergerber, B., Battegay, M., Bernasconi, E., Cavassini, M., Furrer, H., Hoffman, M., Rougemont, M., G\"unthard, H. F. \& Held, L.} (2016).
\newblock {CD4/CD8 ratio and CD8 counts predict CD4 response in HIV-1-infected drug naive and in patients on cART}.
\newblock \textit{Medicine} \textbf{95}, e5094.

\bibitem[{Serrano-Villar(2014)}]{Serrano-Villar:2014}
\textsc{Serrano-Villar, S., Sainz, T., Lee, S. A., Hunt, P. W., Sinclair, E., Shacklett, B. L., Ferre, A. L., Hayes, T. L., Somsouk, M., Hsue, P. Y., Van Natta M. L., Meinert, C. L., Lederman, M. M., Hatano, H., Jain, V., Huang, Y., Hecht, F. M., Martin, J. N., McCune, J. M., Moreno, S. \& Deeks, S. G.} (2014).
\newblock {HIV-infected individuals with low CD4/CD8 ratio despite effective antiretroviral therapy exhibit altered T cell subsets, heightened CD8+ T cell activation, and increased risk of non-AIDS morbidity and mortality}.
\newblock \textit{PLOS Pathogens} \textbf{10}, e1004078.

\bibitem[{Serrano-Villar(2014)}]{Serrano-Villar:2014b}
\textsc{Serrano-Villar, S., Perez-Elias, M.J., Dronda, F., Casado, J. L., Moreno, A., Royuela, A., Perez-Molina, J. A., Sainz, T., Navas, E., Hermida, J. M., Quereda, C. \& Moreno, S.} (2014).
\newblock {Increased risk of serious non-AIDS-related events in HIV-infected subjects on antiretroviral therapy associated with a low CD4/CD8 ratio}.
\newblock \textit{PLOS ONE} \textbf{9}, e85798.

\bibitem[{Shepherd(2016)}]{Shepherd:2016}
\textsc{Shepherd, B. E., Li, C. \& Liu, Q.} (2016).
\newblock {Probability-scale residuals}.
\newblock \textit{Can. J. Stat.} \textbf{44}, 463--479.

\bibitem[{Silva(2018)}]{Silva:2018}
\textsc{Silva, C., Peder, L., Silva, E., Previdelli, I., Pereira, O., Teixeira, J. \& Bertolini, D.} (2018).
\newblock {Impact of HBV and HCV coinfection on CD4 cells among HIV-infected patients: a longitudinal retrospective study}.
\newblock \textit{J. Infect. Dev. Ctries.} \textbf{12}, 1009--1018.


\bibitem[{van der Vaart(1996)}]{vandervaart:1996}
\textsc{van der Vaart, A. W. \& Wellner, J. A.} (1996).
\newblock \textit{{Weak Convergence and Empirical Processes}}.
\newblock Springer.

\bibitem[{Zeng(2006)}]{Zeng:2006}
\textsc{Zeng, D. \& Lin, D. Y.} (2006).
\newblock {Efficient estimation of semiparametric transformation models for counting processes}.
\newblock \textit{Biometrika} \textbf{93}, 627--640.

\bibitem[{Zeng(2007)}]{Zeng:2007}
\textsc{Zeng, D. \& Lin, D. Y.} (2007).
\newblock {Maximum likelihood estimation in semiparametric regression models with censored data}.
\newblock \textit{J. R. Statist. Soc. {\rm B}} \textbf{69}, 507--564.


\end{thebibliography}
\end{document}